\documentclass[]{JFM-FLM_Au}
\usepackage{tabularx}  
\usepackage{textcomp}
\usepackage{upgreek}

\definecolor{viridis0}{rgb}{0.267004, 0.004874, 0.329415}
\definecolor{viridis1}{rgb}{0.283187, 0.125848, 0.444960}
\definecolor{viridis2}{rgb}{0.177423, 0.437527, 0.557565}
\definecolor{viridis3}{rgb}{0.163625, 0.471133, 0.558148}
\definecolor{viridis4}{rgb}{0.214298, 0.355619, 0.551184}
\definecolor{viridis5}{rgb}{0.129933, 0.559582, 0.551864}
\definecolor{viridis6}{rgb}{0.458674, 0.816363, 0.329727}
\definecolor{viridis7}{rgb}{0.993248, 0.906157, 0.143936}

\usepackage{tikz}
\usetikzlibrary{shapes.geometric}

\newcommand{\mDiamond}[1]{%
  \tikz[baseline=-0.6ex]\node[diamond, draw=black, line width=0.5, fill=#1, inner sep=2]{};%
}
\newcommand{\mSquare}[1]{%
  \tikz[baseline=-0.6ex]\node[rectangle, draw=black, line width=0.5, fill=#1, inner sep=3]{};%
}
\newcommand{\mTriUp}[1]{%
  \tikz[baseline=-0.6ex]\node[regular polygon, regular polygon sides=3, shape border rotate=0,
  draw=black, line width=0.5, fill=#1, inner sep=1.5]{};%
}
\newcommand{\mTriDown}[1]{%
  \tikz[baseline=-0.6ex]\node[regular polygon, regular polygon sides=3, shape border rotate=180, draw=black, line width=0.5, fill=#1, inner sep=1.5]{};%
}
\newcommand{\mTriRight}[1]{%
  \tikz[baseline=-0.6ex]\node[regular polygon, regular polygon sides=3, shape border rotate=270,
  draw=black, line width=0.5, fill=#1, inner sep=1.5]{};%
}
\newcommand{\mTriLeft}[1]{%
  \tikz[baseline=-0.6ex]\node[regular polygon, regular polygon sides=3, shape border rotate=90,
  draw=black, line width=0.5, fill=#1, inner sep=1.5]{};%
}

\newcommand{\mDiamondB}[1]{%
  \tikz[baseline=-0.6ex]\node[
    diamond,
    aspect=0.6,          
    draw=black,
    line width=0.5pt,
    fill=#1,
    inner sep=1.5pt
  ]{};%
}

\newcommand{\thickplus}[1]{%
\tikz[baseline=-0.65ex]{
    \draw[black, line width=2.4pt]
        (-0.12,0) -- (0.12,0)
        (0,-0.12) -- (0,0.12);
    \draw[#1, line width=1.7pt]
        (-0.11,0) -- (0.11,0)
        (0,-0.11) -- (0,0.11);
}%
}

\newcommand{\Sone}{\mDiamondB{viridis0}} 
\newcommand{\Stwo}{\mSquare{viridis1}}  
\newcommand{\Sfour}{\mDiamond{viridis2}}
\newcommand{\Mfive}{\thickplus{viridis3}}   
\newcommand{\Lthree}{\mTriDown{viridis2}} 
\newcommand{\Lsix}{\mTriRight{viridis5}}  
\newcommand{\Lseven}{\mTriUp{viridis6}}   
\newcommand{\Leight}{\mTriLeft{viridis7}} 

\lefttitle{M. Bourhis \& O. Buxton}
\righttitle{Journal of Fluid Mechanics}
\usepackage{subcaption}   

\usepackage{hyperref}
\hypersetup{colorlinks = true,urlcolor   = blue, citecolor  = blue}

\title{Dissipation scaling in wind turbine wakes exposed to free-stream turbulence}

\author{Martin Bourhis\aff{1,2}
  \corresp{\email{martin.bourhis@ec-nantes.fr}}
  \and Oliver R.H. Buxton \aff{1}}
 
\affiliation{\aff{1} Department of Aeronautics, Imperial College London, London SW7 2AZ, UK \aff{2}Nantes Université, École Centrale Nantes, CNRS, LHEEA, UMR 6598, F-44 000, Nantes, France}

\corresau{Martin Bourhis, martin.bourhis@ec-nantes.fr}

\begin{document}
\maketitle

\begin{abstract}

The nature of the dissipation of turbulent kinetic energy (TKE) is investigated experimentally in the wake of a diameter $D=0.58\,\mathrm{m}$ wind turbine exposed to several ``flavours'' of high-Reynolds-number free-stream turbulence (FST), produced by an active turbulence-generating grid. For low- and moderate-intensity FST, an annular region of elevated normalised dissipation, $C_{\varepsilon}$, develops in the outer wake, coinciding with a ring of enhanced turbulence intermittency at both small ($\ell \leq \lambda$) and large ($\ell \geq D$) scales, where $\lambda$ is the Taylor microscale. In the blade-tip region, $C_{\varepsilon}$ scales with $\sqrt{Re_D}/Re_{\lambda}$, where $Re_D$ is a global Reynolds number and $Re_{\lambda}$ a local turbulent Reynolds number based on $\lambda$---a scaling indicative of dissipation being out of equilibrium with the inter-scale flux of TKE in the inertial range of the energy cascade. This non-equilibrium regime is interpreted in light of the observed intermittency : large-scale intermittent events (i.e., low-wavenumber perturbations/``kicks''), driven by persistent tip-vortex and tip-shear-layer dynamics under low-intensity FST, require a finite time to cascade down to the dissipative scales, thereby introducing an imbalance between the inter-scale energy flux and dissipation. At the wake centreline, by contrast, $C_{\varepsilon}$ remains approximately constant with streamwise distance, reflecting either classical Kolmogorov-type equilibrium turbulence or balanced non-equilibrium turbulence, with intermittency confined to the small scales. Under high-intensity FST, large-scale intermittency is suppressed, consistent with the erosion of tip-vortex structures, and no comparable scaling for $C_{\varepsilon}$ could be identified using a single turbulent Reynolds number for these cases, where two similarly intense streams of turbulence, but of different origins, are adjacent to one another.

\end{abstract}


\newpage

\section{Introduction \label{sec:Introduction}}

Wind energy has emerged as a cornerstone of the global energy transition, with 165 GW of new capacity installed in 2025, a 40\% increase relative to 2024 \citep{gwec2026}. 
Wind turbines operate within the atmospheric boundary layer (ABL) and are commonly deployed in densely packed arrays, known as wind farms, to maximise the areal power density extracted from a finite lease area whilst sharing infrastructure costs. 
Such clustering inevitably produces turbine-wake interactions: the turbulent wakes shed by upstream rotors interact with the ambient atmospheric turbulence to form the unsteady inflow impinging on downstream machines.
The resulting ``wake effects'' on wind-farm efficiency are twofold: wake-induced momentum deficits reduce overall farm energy production \citep{Kirby2025,Barthelmie2025}, whilst wake-added turbulence amplifies unsteady blade aerodynamic loading, accelerating fatigue damage and shortening turbine lifetime \citep{Pacheco2024,Xu2025,Oliveira2025}.
Predicting and ultimately mitigating these power losses through wind-farm layout optimisation and control is therefore a central challenge in wind-energy science \citep{Veers2019}, requiring both a thorough fundamental understanding of turbulent wake physics and computationally efficient models that capture key wind turbine wake statistics \citep{Meneveau2019}.

This challenge has motivated two decades of sustained research---spanning wind-tunnel experiments, field measurements, and high-fidelity simulations---aimed at characterising the multi-scale interactions between the atmosphere and wind-energy  systems. Wind turbines generate complex turbulent flows comprising multiple coherent structures : tip and root vortices, trailing-edge vortex sheets, vortex shedding from the nacelle and tower, and large-scale wake meandering \citep[e.g.][]{Stevens2017,PorteAgel2019,Biswas2024}. 
The near-field flow is three-dimensional, swirling, and strongly inhomogeneous, with the wake structure retaining the imprint of the turbine components (e.g. the blades, nacelle, tower) \citep{Vermeer2003,Cillis2021,Dong2023}. 
Beyond tip-vortex breakdown, conventionally taken to mark the onset of the far field, the wake exhibits more universal behaviour : the rotor's influence on the mean flow is often approximated by that of a simple momentum-extracting porous body, encapsulated in a global aerodynamic parameter, the thrust (or drag) coefficient \citep{Sorensen2016}. The transition from the near to the far field, and more broadly wake recovery, is strongly affected by the surrounding atmospheric turbulence. Both the intensity and the characteristic length and time scales of the free-stream turbulence (FST) have been shown to influence wind-turbine wake dynamics and mean-flow statistics \citep{Gambuzza_Ganapathisubramani_2023, Hodgson2023, Bourhis2025, Biswas_Buxton_2026}. In addition, an important feature of wind-turbine wakes is the presence of an annular region of strong turbulence intermittency (extreme deviation from the mean fluctuation level) surrounding the mean velocity deficit \citep[e.g.][]{Schottler2018,Neunaber2020}. The flow in this intermittency ring exhibits non-Gaussian, heavy-tailed velocity-increment distributions at scales larger than the turbine diameter, whereas the wake core exhibits intermittency only at dissipative scales, as in canonical homogeneous turbulence. Whether turbine-generated or inherited from background atmospheric turbulence, intermittency affects wind-farm performance in two ways: it increases torque and power fluctuations \citep{Mucke2011,Chamorro2015_Intermittency,Schottler2017}, and it drives sporadic, high-amplitude loading events that accelerate blade fatigue \citep{Schwarz2019,Wang2025}. The intermittency ring was first characterised under low-turbulence inflow conditions \citep{Schottler2018} and later shown to persist in turbulent ambient flows \citep{Kadum2019,Zheng2023}; however, its dependence on FST intensity, FST integral length scale, and the turbine operating point remains largely unexplored.

Due to the complexity of wind turbine wake turbulence, most analytical models have focused on bulk statistics, notably the evolution of the mean velocity deficit. These models typically assume a self-similar, axisymmetric far-field mean flow with Gaussian velocity deficit radial profiles, and empirical spreading rates fitted to experimental, numerical, or field data \citep[see][and references therein]{Stevens2017,PorteAgel2019}. Although valuable for engineering wind-farm design, these models rely heavily on empirical coefficients and lack direct grounding in classical turbulence theory. However, from a fundamental standpoint, wind-turbine wakes are, to first order, related to an extensively studied canonical free shear flow : the axisymmetric turbulent wake \citep[for a historical review, see][]{Johansson2003}. In particular, \citet{Townsend1976} and \citet{George1989} derived an axial scaling of the maximum (centreline) mean-velocity deficit ($\upDelta u_{0}$) with downstream distance ($x$) for high-Reynolds-number flows past axisymmetric solid (i.e. non-porous) bodies developing in a quiescent ambient flow : $\upDelta u_{0} \sim (x - x_0)^{-2/3}$, where $x_0$ is a virtual origin that emerges naturally from Townsend-George equilibrium similarity analysis. This scaling relies on two hypotheses: (i) self-preservation of one-point turbulence statistics (e.g. mean flow, Reynolds stresses, turbulent kinetic energy) as well as the dissipation rate of turbulent kinetic energy (a one-point statistic that is in fact the limit of $r\rightarrow 0$ for a two-point statistic), and (ii) a closure on the self-similar form of the dissipation rate ($\varepsilon$) of turbulent kinetic energy (TKE, $K$) within the wake : $\varepsilon = C_{\varepsilon} {K^{3/2}}/{\mathcal L}$, where the normalised dissipation rate $C_{\varepsilon}$ is constant (the Taylor-Kolmogorov dissipation scaling). Here, ${\cal L}$ is an integral length scale of the turbulence, characterising the size of the largest (energy-containing) eddies in the wake; a two-point statistic. The $-2/3$ power law decay is commonly referred to as the ``equilibrium" scaling law, in reference to the Richardson–Kolmogorov statistically steady equilibrium TKE cascade, where the dissipation rate at the smallest scales balances instantaneously the inter-scale energy flux across inertial-range scales, yielding $C_{\varepsilon} = \textrm{const.}$

Over the past decade, however, substantial evidence has accumulated showing that the normalised dissipation rate does not always follow the classical scaling $C_{\varepsilon} \approx \textrm{const.}$, but can instead scale with the ratio of two Reynolds numbers : a global Reynolds number $Re_G$ (defined by the inlet, initial or boundary conditions), and a local turbulent Reynolds number, which varies in space and time \citep[see][for a review, and references therein]{Vassilicos2015}. In particular, the scaling $C_{\varepsilon}~\sim Re_G^{m/2}/Re_{\lambda}^n$, where $m\approx1$, $n\approx1$, and $Re_{\lambda}$ is the Taylor length-based Reynolds number, has been reported both experimentally and numerically across a wide range of turbulent flows, including fractal and regular grid-generated decaying turbulence \citep{Seoud_2007,Valente_2011,Valente2012,Valente2015}, axisymmetric and planar wakes of multi-scale (fractal) bluff bodies \citep{Nedic2013,Dairay2015,Obligado2016,Noriega_2025}, unsteady (decaying and forced) periodic turbulence \citep{Goto2015,Goto2016a,Goto2016b}, interacting wakes of side-by-side square prisms \citep{Chen_Cuvier_Foucaut_Ostovan_Vassilicos_2021}, atmospheric flows \citep{Waclawczyk2022,Waclawczyk2022_2}, zero-pressure-gradient turbulent boundary layers \citep{Nedic_2017}, and fully-developed turbulent channel flows \citep{Apostolidis_Laval_Vassilicos_2022}. This scaling for $C_{\varepsilon}$ violates the classical Richardson-Kolmogorov (equilibrium) phenomenology for the cascade of TKE, and is hence referred to as a ``non-equilibrium" dissipation scaling \citep{Vassilicos2015}. Incorporating this non-equilibrium scaling, $C_{\varepsilon}\sim\sqrt{Re_G}/Re_{\lambda}$, within the self-similar analysis of \citet{George1989}, \citet{Dairay2015} derived a new power law for the streamwise evolution of the centreline mean velocity deficit, $\upDelta u_0~\sim(x-x_0)^{-1}$. This $-1$ power law decay has since been reported in multi-scale bluff-body axisymmetric wakes \citep{Nedic2013,Dairay2015}, in the near wake of porous discs designed to match wind-turbine wake aerodynamics \citep{Lingkan2023,Bourhis2024,Noriega_2025}, and in laboratory- and full-scale wind-turbine wakes \citep{Neunaber2020,Neunaber2022,Neunaber2024}. An alternative power-law decay, $\upDelta u_0 \sim (x-x_0)^{-2}$, has also been reported for both porous-disc \citep{Bourhis2024} and wind-turbine models \citep{Stein2019}, and may be interpreted in terms of non-equilibrium scaling with $m \approx n \approx 2$, although the dissipation rate was not directly quantified in these studies. It is worth noting that \citet{Goto2016b} also showed that $C_{\varepsilon}=\textrm{const.}$ does not necessarily imply a classical Kolmogorov-type equilibrium cascade, but may also arise under ``balanced'' non-equilibrium conditions, in which the cascade remains strongly non-stationary but the inter-scale energy flux and dissipation remain approximately proportional to one another. Taken together, these results suggest that wind-turbine wake turbulence may depart from classical equilibrium dissipation, and that the George–Townsend similarity analysis could provide a theoretical framework for wind-turbine wake modelling, provided that the nature of the dissipation is known throughout the wake and across the range of inflow and operating conditions encountered by a turbine.

Much of the previous work, however, has considered two simplifying configurations: (i) solid (non-porous) wake generators, and (ii) wakes developing within a non-turbulent free stream. This simplifies the problem since there is only one ``flavour'' of turbulence within such a flow---that generated within the wake itself arising from mean shear---and not an interaction between two adjacent, yet fundamentally different streams of turbulence. In addition, solid wake-generators exclude any flow through the wake generator. Neither condition is representative of an operating wind turbine: the rotor plane is porous, meaning that there is a flow through it, and, since turbines are situated within the turbulent ABL, their wakes are exposed to background turbulence whose history is completely different to that of the wake turbulence. Both FST and the intrinsic characteristics of wind-turbine wakes (intermittency ring, multi-scale coherent structures) may modify the energy cascade and the nature of the dissipation, with potentially different behaviours between the wake core and the dynamically active and intermittent outer edge \citep{Schmitt2025}. For instance, \citet{Bourhis2024} reported different scalings for $\upDelta u_0(x)$ depending on the characteristics of the ambient turbulence and the disc porosity. Finally, \citet{Rind2012} showed that sufficiently strong external turbulence can suppress the self-similarity of wakes in the far field, which would preclude the application of the George–Townsend similarity analysis in the presence of FST.

Therefore, the goal of the present study is to investigate the nature of turbulence dissipation in wind-turbine wakes exposed to several different ``flavours'' of high-Reynolds-number free-stream turbulence. In particular, we seek to assess whether the wake turbulence follows an equilibrium or a non-equilibrium dissipation scaling, and how the dissipation varies with the turbine operating point, with the intensity and integral length scale of the FST, and spatially across the wake (e.g. between the wake core and the intermittent outer edge). The paper is organised as follows. \S~\ref{sec:Experimental setup} describes the experimental set-up and the FST parameter space examined. \S~\ref{sec:mean_flow_and_turb_fields} provides an overview of the influence of FST on the mean flow and turbulence fields. The self-similarity of one-point turbulence statistics is then examined in \S~\ref{sec:self_sim}, the scaling of the turbulent kinetic energy dissipation rate in \S~\ref{sec:Scaling of the time-averaged turbulent energy dissipation rate}, and wake intermittency in \S~\ref{sec:wake intermittency}.

\section{Methodology and inflow parameter space \label{sec:Experimental setup}}

\begin{figure}
    \begin{subfigure}[c]{0.72\linewidth}
        \emph{(a)} \par \vspace{0.1cm}
        \centering
        \includegraphics[width=\linewidth]{Figures/ExperimentalSetup/Setup_5.pdf}
    \end{subfigure}
    \hfill
    \begin{subfigure}[c]{0.22\linewidth}        
        \emph{(b)} \par \vspace{0.1cm}
        \centering
        \includegraphics[width=\linewidth]{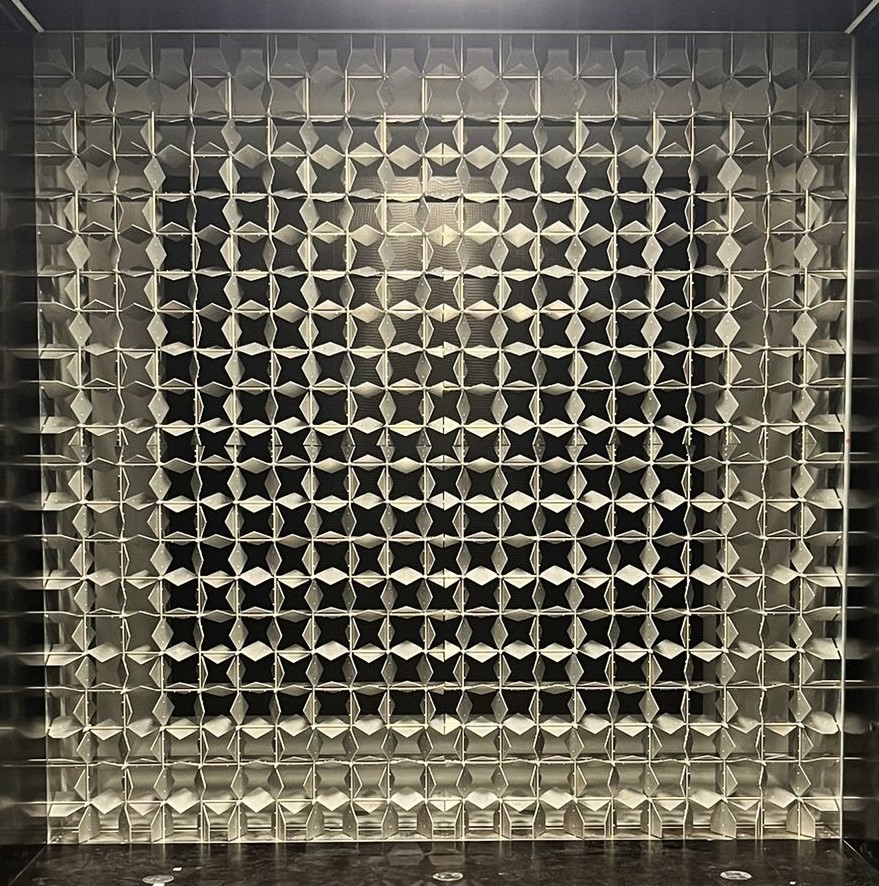} \\[0.1cm]
        \includegraphics[width=\linewidth]{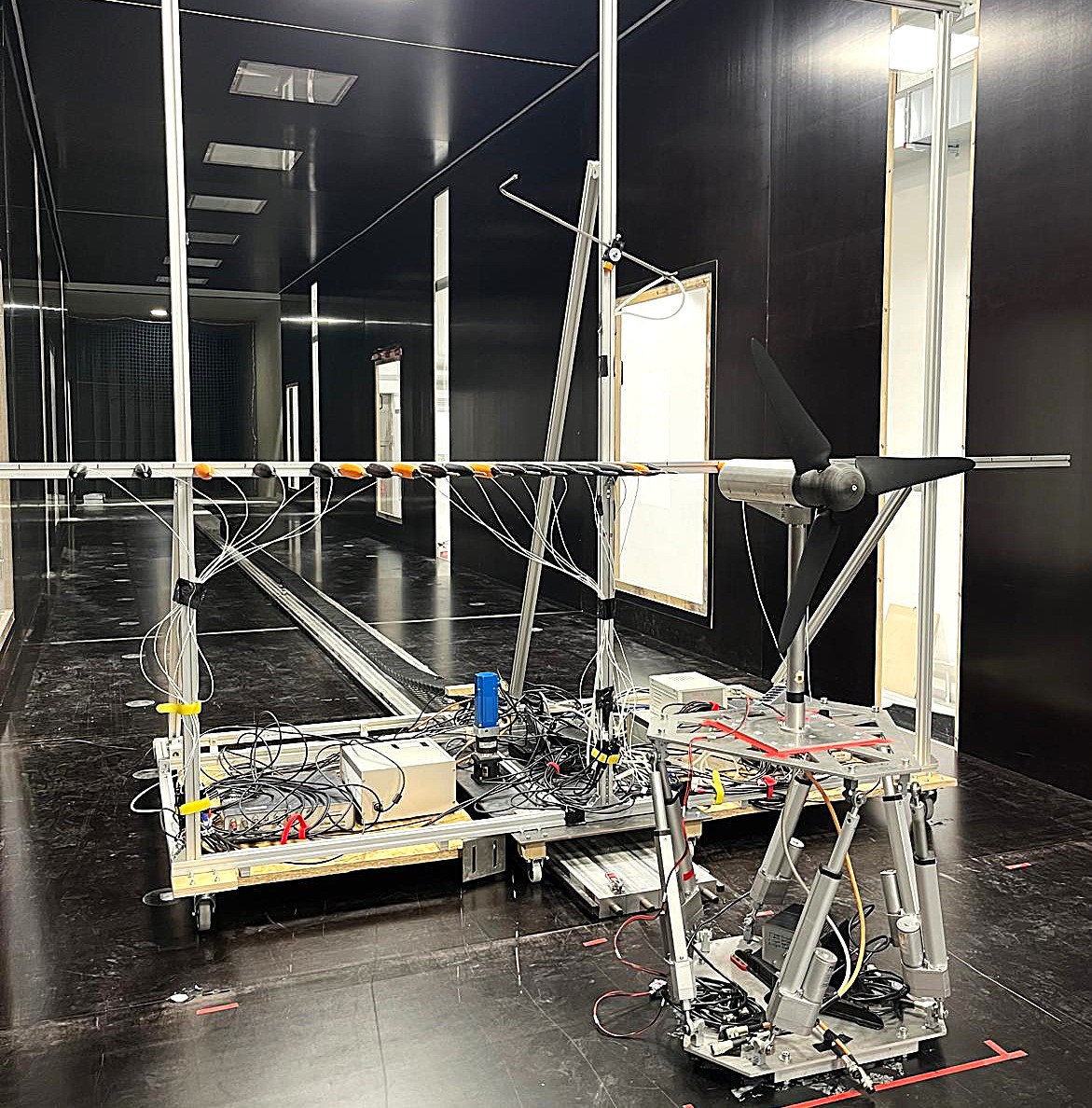}
    \end{subfigure}
    \caption{Schematic and photographs of the experimental setup.}
    \label{fig:setup}
\end{figure}

The dataset used in this study was obtained from a wind tunnel campaign by \citet{Bourhis2025} in which wind turbine wake measurements were performed for various operating and inflow conditions. Therefore, only the main experimental parameters are summarised here, and the reader is referred to the aforementioned study for further details.

\subsection{Experimental setup}

Experiments were conducted in the $3~\textrm{m} \times3~\textrm{m} \times30~\textrm{m}$ test section of the closed-loop wind tunnel at the University of Oldenburg. A three-bladed wind turbine model, with a rotor diameter of $D = 0.58~\text{m}$, was exposed to eight customised turbulent inflow conditions produced by an active turbulence-generating grid installed at the outlet of the contraction nozzle (figure~\ref{fig:setup}). The time-averaged free-stream velocity was kept fixed at $U_{\infty} = 6.5~\text{m.s}^{-1}$, yielding a diameter-based Reynolds number $Re_{D} = 2.55 \times 10^5$. For each turbulent inflow, the turbine was operated at three tip-speed ratios, $\textrm{TSR} \in \{1.7, 2.7, 3.7\}$ corresponding to low ($C_T \approx 0.5$), medium ($C_T \approx 0.7$), and high ($C_T \approx 0.9$) thrust (drag) coefficients. The tip-speed ratio is defined as the ratio of the blade tip velocity to the incoming free-stream velocity, $\textrm{TSR} = D\omega/2U_{\infty}$, where $\omega$ denotes the angular velocity of the turbine. The blade pitch angle was held fixed across all operating conditions.

For all 24 \{$C_T$, FST\} combinations, hot-wire anemometry measurements were conducted downstream of the wake generator. Time-resolved streamwise velocity signals, $u(x,y,t)$, were acquired over $x/D\in [1,20]$, at increments of $1D$, using a linear arrangement of 21 single hot-wire probes distributed along the horizontal direction ($y$) at hub height ($z = 0$) and mounted on a motorised traverse system. Figure~\ref{fig:setup} shows the streamwise and spanwise measurement stations; hot-wire probes shown in red ({\color{red} $\blacksquare$}) were operated with a Dantec Dynamics StreamLine 9091N0102 frame with 91C10 CTA modules and sampled at 20~kHz, whereas those in blue ({\color{blue} $\blacksquare$}) used 54N80 CTA modules and were sampled at 6~kHz. Most probes were positioned on one side of the turbine, with an increased density near the geometric centreline ($y/D = 0$). Turbine alignment with the incoming flow was checked prior to the measurements, and wake symmetry was assessed using hot-wire probes on the negative $y$ side. Velocity signals were acquired for 240~s at each measurement station, ensuring statistical convergence of all turbulent quantities presented in this work. Hot-wire probes were calibrated at the beginning and end of each experimental day with the test section empty. The reference velocity, $U_{\infty} = \sqrt{2{\upDelta} p}/\rho$, was obtained from the differential pressure $\upDelta p$ measured by a Prandtl tube placed in the free stream. The air density $\rho$ was corrected for ambient temperature, relative humidity, and atmospheric pressure following \cite{Davis_1992}. Hot-wire voltage signals ($V$) were temperature-corrected following \cite{Bruun_1995}. Calibration coefficients were determined by fitting a fourth-order polynomial to the \{$V$, $U_{\infty}$\} data. To account for slow hot-wire drift associated with ambient variations over the duration of the experimental day, the velocity signals were reconstructed using a linear time-weighted combination of the two bounding calibration laws.

\subsection{Inflow turbulence statistics and signal processing}

\begin{table}
\centering
\setlength{\tabcolsep}{7pt} 
\def~{\hphantom{0}}
  \begin{tabular}{lcccccccc}
       \multicolumn{2}{c}{FST Case} &  Mode  &   $I_{\infty} (\%)$  & $Re_{\lambda,\infty}$ & $\lambda_{\infty}$~(mm) & $\eta_{\infty}$~(mm) &  ${\cal L}_{\infty}$~(cm) &  ${\cal L}_{\infty}/D$  \\[4pt]
       S1 & \Sone   & - & ~0.9 (0.1) & ~~55 ~~(5) & 13.8 (0.9) & 0.97 (0.05) & ~~4.9 ~(0.5) & 0.1  \\ [1pt] 
       S2 & \Stwo   & SM & ~1.8 (0.1) & ~100 ~~(5) & 14.4 (0.7) & 0.73 (0.03) & ~~9.2 ~(0.8) & 0.2 \\ [1pt]
       S4 & \Sfour  & SM & ~4.5 (0.4) & ~280 ~(40) & 14.4 (0.7) & 0.44 (0.02) & ~20.2 ~(2.1) & 0.3 \\ [1pt]
       M5 & \Mfive  & DM & ~4.9 (0.3) & ~430 ~(55) & 22.4 (1.9) & 0.55 (0.02) & ~67.7 (14.8) & 1.2 \\ [1pt]
       L3 & \Lthree & DM & ~3.7 (0.3) & ~385 ~(60) & 25.6 (2.3) & 0.66 (0.02) & 102.1 ~(9.9) & 1.8\\ [1pt]
       L6 & \Lsix   & DM & ~5.8 (0.4) & ~695 (100) & 29.2 (2.5) & 0.56 (0.02) & 112.8 (11.4) & 1.9 \\ [1pt]
       L7 & \Lseven & DM & ~8.8 (0.4) & ~935 (120) & 25.6 (2.1) & 0.43 (0.01) & 113.3 (15.8) & 2.0 \\ [1pt]
       L8 & \Leight & DM & 10.8 (0.5) & 1060 (120) & 23.5 (1.5) & 0.37 (0.01) & 116.8 (16.8) & 2.0\\    
  \end{tabular}
  \caption{Summary of the inflow turbulence characteristics for all experimental cases. The table reports the active-grid operating mode (SM for static mode and DM for dynamic mode), the free-stream turbulence intensity $I_{\infty}$, the Reynolds number based on the Taylor microscale $Re_{\lambda,\infty}$, the Taylor microscale $\lambda_{\infty}$, the Kolmogorov microscale $\eta_{\infty}$, and the streamwise integral length scale ${\cal L}_{\infty}$. Reported values correspond to the average across the tunnel width ($y (\textrm{m}) \in [-1.1,1.1]$), with the standard deviation in parentheses.}
  \label{tab:FST}
\end{table}

\begin{figure}
    \begin{subfigure}[c]{0.47\linewidth}
        \emph{(a)} \par \vspace{0.1cm}
        \centering
        \includegraphics[width=\linewidth]{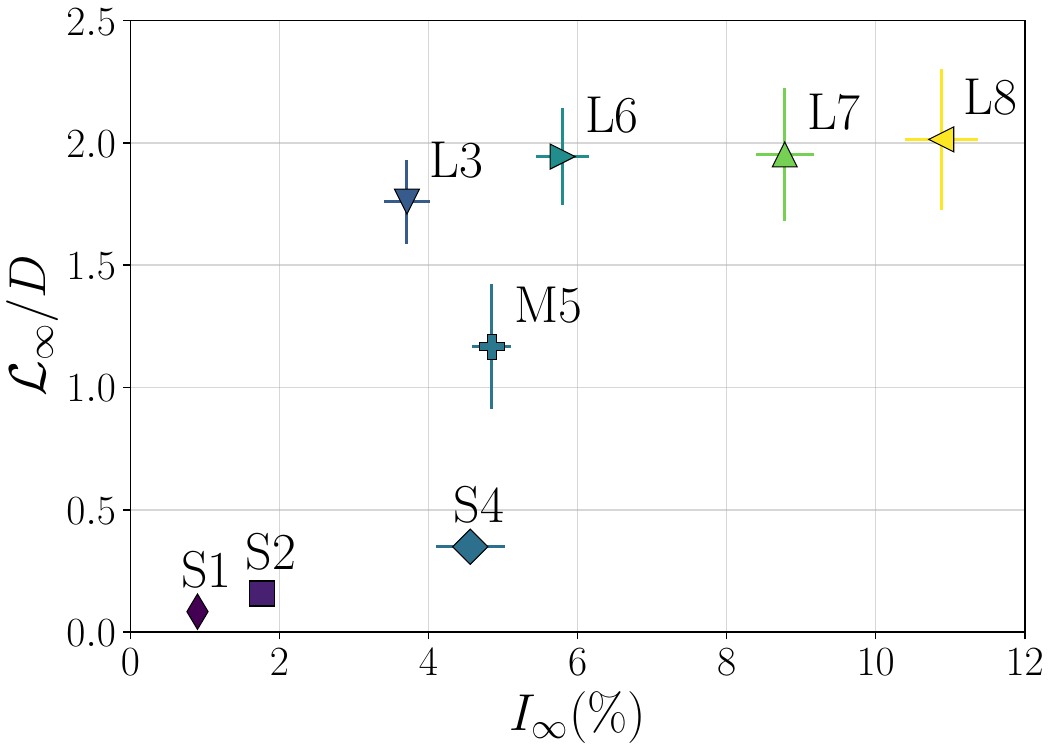}
    \end{subfigure}
    \hfill
    \begin{subfigure}[c]{0.47\linewidth}        
        \emph{(b)} \par \vspace{0.1cm}
        \centering
        \includegraphics[width=\linewidth]{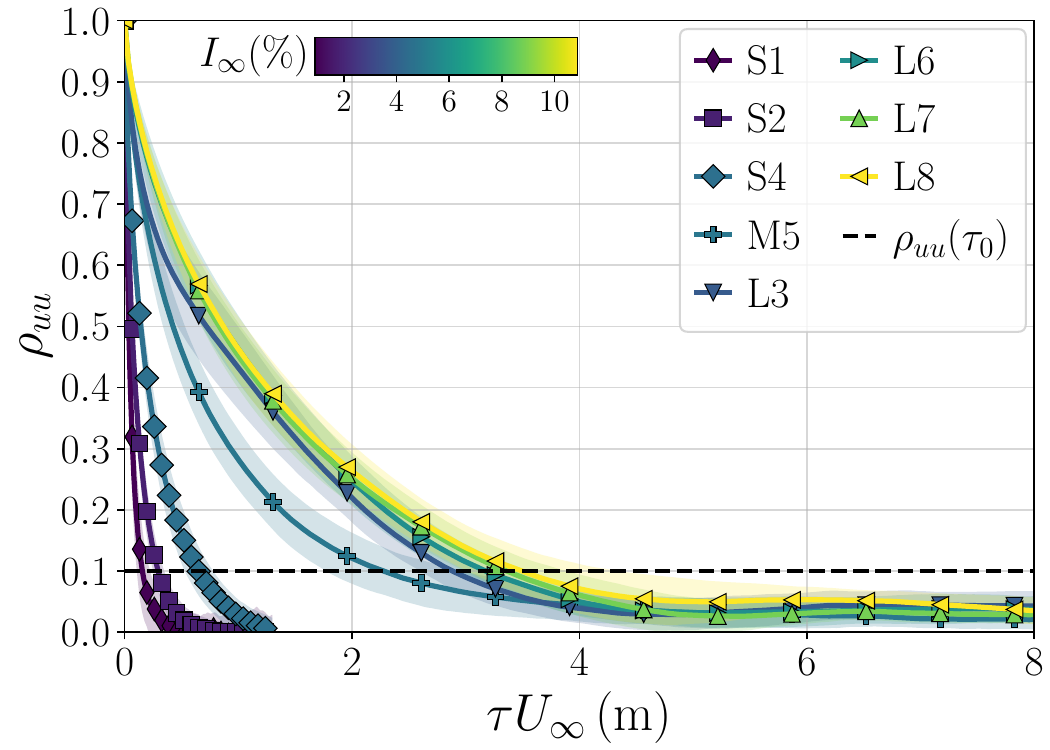}
    \end{subfigure}
    \\
        \begin{subfigure}[c]{0.47\linewidth}        
        \emph{(c)} \par \vspace{0.1cm}
        \centering
        \includegraphics[width=\linewidth]{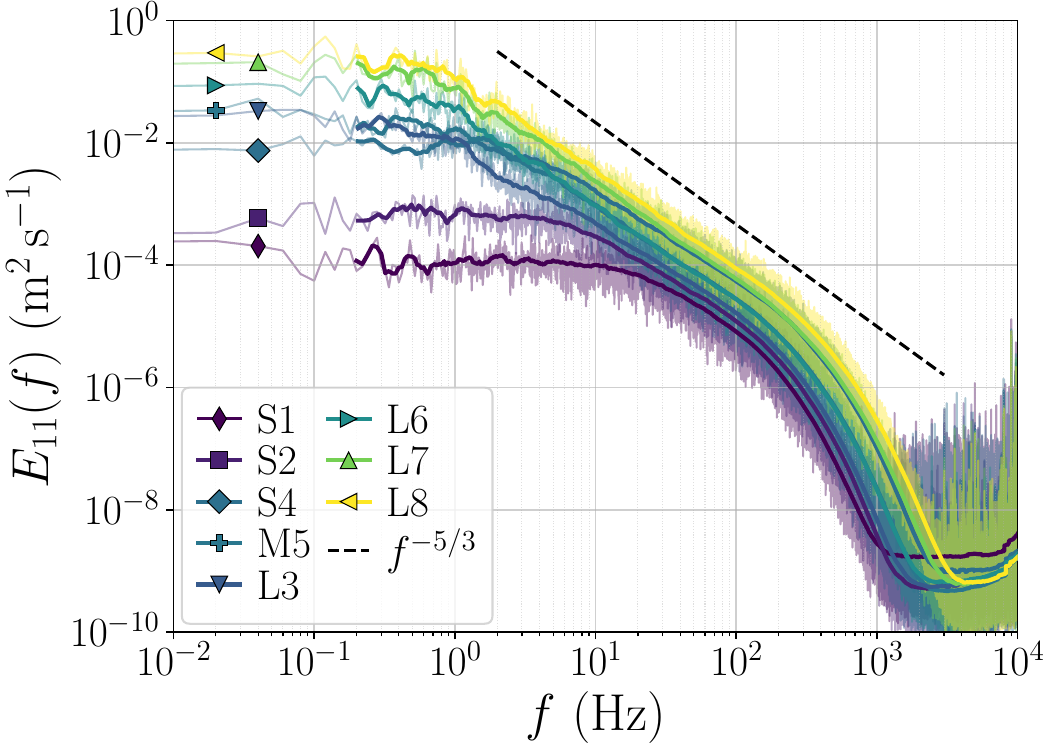}
    \end{subfigure}
     \hfill
    \begin{subfigure}[c]{0.47\linewidth}        
        \emph{(d)} \par \vspace{0.1cm}
        \centering
        \includegraphics[width=\linewidth]{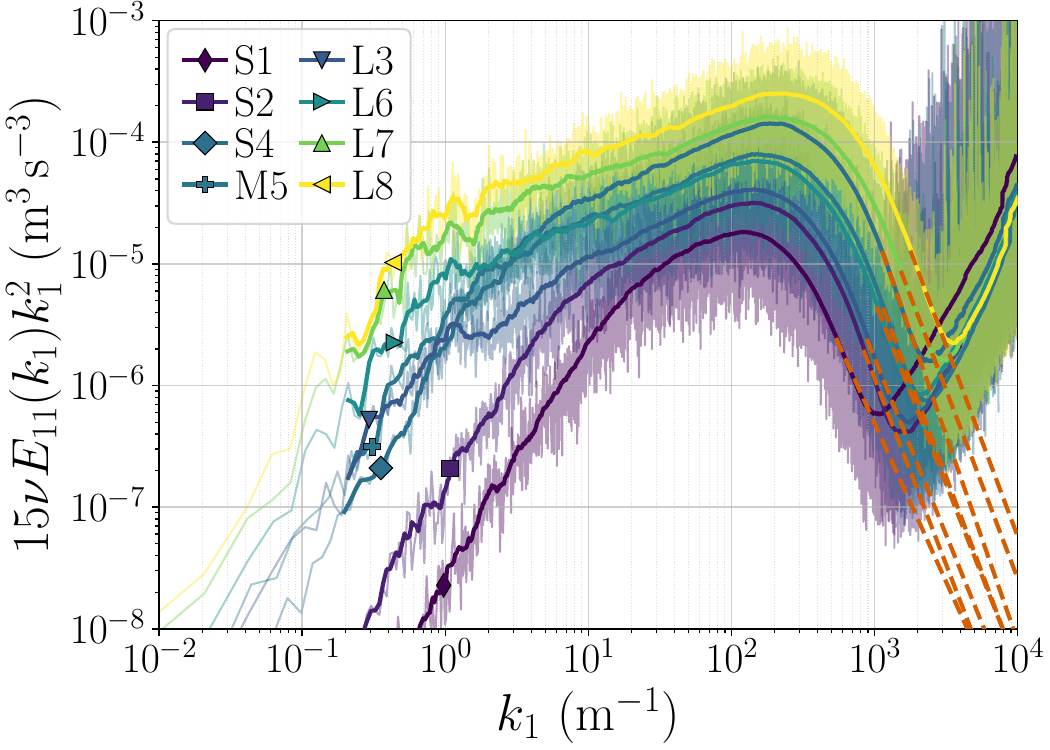}
    \end{subfigure}
    \caption{\emph{(a)} Free-stream turbulence parameter space $\{I_{\infty}(\%),{\cal L}_{\infty}/D \}$. Each point corresponds to the mean across the tunnel width ($y (\textrm{m}) \in [-1.1,1.1]; x=0;z=0$); error bars indicate one standard deviation. (\emph{b}) Normalised streamwise velocity autocorrelation functions, $\rho_{uu}$ averaged across the tunnel width; the shaded region denotes one standard deviation. The dashed line marks the threshold $\rho_{uu}(\tau_0) =0.1$ used to define ${\cal L}_{\infty}$. \emph{(c)} FST power spectral densities of velocity fluctuations measured at the hub location. \emph{(d)} FST premultiplied spectra. Dashed orange lines indicate power-law fits of the noisy high-frequency region used to estimate the dissipation rate $\varepsilon$. }
    \label{fig:fst}
\end{figure}

Eight user-specified turbulent inflows were generated using the $3~\textrm{m} \times3~\textrm{m}$ active-turbulence generating grid. The grid consists of 80 independently actuated rotating shafts (40 horizontal and 40 vertical), each spanning half the grid and fitted with rectangular flaps, resulting in a mesh size of $M = 14.3~\textrm{cm}$. Each shaft is actuated by a servo motor, allowing its rotation rate, direction, and amplitude to be varied dynamically. The eight turbulent inflows were produced using two grid actuation modes with distinct protocols. In the dynamic mode (DM), the flaps oscillate back and forth according to predefined shaft-angle time series, whereas in the static mode (SM), the shafts are fixed at prescribed angles. Further details of the protocols are provided in Appendix~\ref{appA}, and movies of the grid in operation are available in \citet{Bourhis2025}. Prior to the wake measurements, all inflow conditions were characterised with the turbine removed from the wind tunnel. Hot-wire probes were evenly distributed across the wind-tunnel width, $y (\textrm{m}) \in [-1.4,1.4]$, in the turbine swept plane ($\{x;z\} =\{0;0\}$). The inflow turbulence characteristics---turbulence intensity, fundamental turbulence length scales, and spectral content---are summarised in table~\ref{tab:FST} and figure~\ref{fig:fst}. 

The streamwise velocity was decomposed as $u(x,y,t) =  U(x,y) + u'(x,y,t)$, where $U(x,y) = \langle u(x,y,t) \rangle$ denotes the time-averaged velocity and $u'(x,y,t)$ the velocity fluctuations. The turbulence intensity was defined as $I= \sqrt{\langle u^{'2}\rangle }/U$. The integral length scale $\mathcal{L}$ was estimated from the normalised autocorrelation function of the temporal velocity fluctuations, $\rho_{uu}$, as ${\cal L} = U \int_0^{\tau_0}\rho_{uu}(\tau)d\tau$, where $\tau_0$ is the time lag such that $\rho_{uu}(\tau_0) = 0.1$ (figure~\ref{fig:fst}(\emph{b})). The conversion from temporal to spatial scales relies on Taylor’s frozen-turbulence hypothesis, which is justified by the low turbulence intensity (always below, and in most cases much smaller than, 11\%). The influence of the integration limit used to estimate the integral length scale $\mathcal{L}$ was assessed. Larger integration limits (e.g. $1/\mathrm{e}$ \citep{Tritton1977,Fuchs2022}), yielded smaller values of $\mathcal{L}$ with reduced spatial dispersion between probes, whereas smaller limits produced larger values with increased dispersion (integration up to the first zero crossing was not considered, as the autocorrelation function does not consistently cross zero for some DM protocols). The bias associated with the choice of $\tau_0$ was assessed by repeating the entire analysis presented in this work using the $1/\mathrm{e}$ criterion. Although the absolute values of $\mathcal{L}$ varied, the reported trends and conclusions remained unchanged.


The mean turbulent-kinetic-energy dissipation rate, $\varepsilon$, was estimated by integrating the dissipation spectrum, $\varepsilon = \int_0^{k_{max}}k_1^{2} 15 \nu E_{11}(k_1) dk_1$ (figure~\ref{fig:fst}(\emph{d})). Here, $E_{11}(k_1) = E_{11}(f)U/2\pi$ is the one-dimensional power spectral density, $k_1 = 2\pi f/U$ the wavenumber, $f$ the Fourier frequency, and $\nu=1.5\times 10^{-5}~\textrm{m}^{2}\,\textrm{s}^{-1}$ the kinematic viscosity. This formulation relies on local small-scale isotropy and, again, on the Taylor hypothesis. Although local isotropy was not assessed directly, the measurements were performed sufficiently far downstream of the grid (13.4~m, $\approx94M$) to justify this assumption based on previous active-grid studies \citep{Thormann2014,Hearst2015,Mora2019}. The longitudinal power spectral densities, $E_{11}(f)$, shown in figure~\ref{fig:fst}(\emph{c}) were estimated via a fast-Fourier-transform-based periodogram algorithm using a Hann window with 50\% overlap and a 50~s window length, corresponding to at least 270 integral length scales for the cases L\# and substantially more for all other FST cases.

Following \citet{Mora2019} and \citet{Fuchs2022}, the high-wavenumber tail of the dissipation range was modelled using a power-law fit applied to each time signal to reduce unavoidable high-frequency noise contamination, which can bias the estimate of $\varepsilon$. The dissipation rate was therefore obtained by integrating the measured spectrum from $k_1=0$ to the lower bound of the fitted wavenumber range, and the modelled spectrum from this bound to the maximum resolved wavenumber ($k_{max}$). For all measurements, we checked that the main contribution to the integral comes from the measured dissipation spectrum, with the power-law-modelled high-wavenumber contribution remaining below 7\% of the estimated value of $\varepsilon$. The Taylor microscale $\lambda$, Kolmogorov microscale $\eta$, and Taylor-scale Reynolds number $Re_{\lambda}$ were subsequently estimated from $\varepsilon$ as $\lambda=\sqrt{15\nu\langle u^{'2}\rangle/\varepsilon}$, $\eta=(\nu^3/\varepsilon)^{1/4}$, $Re_{\lambda} = \sqrt{\langle u^{'2}\rangle}\lambda/\nu$. The free-stream turbulence envelopes (denoted by the subscript $\infty$) for all inflows, $\{I_{\infty},Re_{\lambda,\infty}, \lambda_{\infty} ,\eta_{\infty}, {\cal L}_{\infty} \}$, shown in table~\ref{tab:FST} and figure~\ref{fig:fst}, were obtained by spatial averaging across the hot-wire probe array; e.g. $I_{\infty} = \langle I \rangle_{y(\mathrm{m})\in [-1.1,1.1]}$, with the same procedure applied to all other quantities. 

The resulting inflow conditions span a broad range of turbulence regimes, with FST intensity varying over $I_{\infty} (\%) \in [1\%,11\%]$, integral length scales over ${\cal L}_{\infty}/D \in [0.1,2]$, and more than one order of magnitude variation in the Taylor-scale Reynolds number $Re_{\lambda,\infty} \in [55,1060]$. The grid protocols were tailored to generate inflows with specific ratios of the turbulence integral length scale to the turbine diameter and were accordingly classified as cases \textbf{S} (small, ${\cal L}_{\infty}<D/2$), \textbf{M} (medium, ${\cal L}_{\infty}\sim D$), and \textbf{L} (large, ${\cal L}_{\infty}\sim2D$). The cases were then numbered hierarchically by increasing turbulence intensity, $I_{\infty}$, from \#1 to \#8, with the eight FST ``flavours'' $\{ I_{\infty}, {\cal L}_{\infty}\}$ graphically presented in figure~\ref{fig:fst}(\emph{a}). 

In case S1, the grid is fully opened, yielding an inflow with a low turbulence intensity ($I_{\infty} = 0.9\%$), a small integral length scale (${\cal L}_{\infty}/D\approx 0.1$), and a low Taylor-scale Reynolds number ($Re_{\lambda,\infty} = 55$). Consistently, the corresponding power spectrum exhibits weak low-wavenumber energy content and a very narrow inertial range (less than half a decade, figure~\ref{fig:fst}(\emph{c})). S1 therefore provides the closest approximation to a non-turbulent background flow and is used as the baseline case. The two static-mode cases (S2 and S4) produce turbulent inflows with characteristics comparable to traditional regular passive-grid-generated turbulence, with moderate Taylor-scale Reynolds numbers and turbulence intensities, and relatively small integral length scales \citep{Valente_2011,Valente2012,Valente2015}. 

As expected, turbulent inflows generated by the active grid in dynamic mode exhibit larger Taylor-scale Reynolds numbers, broader inertial ranges, larger integral length scales, and generally higher turbulence intensities \citep{Mydlarski_Warhaft_1996,Thormann2014,Shet2020}. The low-frequency forcing introduced by the grid motions (see excitation spectra in Appendix~\ref{appA}) injects energy at the largest scales, which is subsequently transferred towards smaller scales as the flow develops between the grid and the turbine location. All DM-inflows exhibit a Kolmogorov-like spectral distribution with a clear $\mbox{-5/3}$ slope over the inertial range and no spectral gap between the production and inertial ranges, indicating that the turbulence is fully developed at the measurement location. Extended inertial subranges and high Taylor-scale Reynolds numbers are observed for all DM cases, with the most turbulent inflow, L8, reaching $Re_{\lambda,\infty}=1060$ and exhibiting a $-5/3$ spectral decay over $\approx 2$~decades.

By adjusting the shaft-angle standard deviation and rotation rate (see Appendix~\ref{appA}), both the intensity and the integral length scale of the background turbulence were varied quasi-independently. The L\# cases primarily increase turbulence intensity at approximately constant integral length scale, whereas the cases \{L3, S4, M5, L6\} exhibit, \emph{a priori}, a more pronounced increase in the turbulence integral length scale relative to intensity. This FST parameter space therefore enables the separate quantification of their respective influences on wake development. Finally, all inflow conditions were verified to exhibit horizontally uniform first- and second-order turbulence statistics, together with negligible decay of the turbulence intensity over the wake measurement region (with a maximum variation of $\upDelta I_{\infty} = 0.02$ for case L8).

\begin{figure}
    \begin{subfigure}[c]{\linewidth}
    \emph{(a)} \par \vspace{0.1cm}
            \centering
        \includegraphics[width=\linewidth]{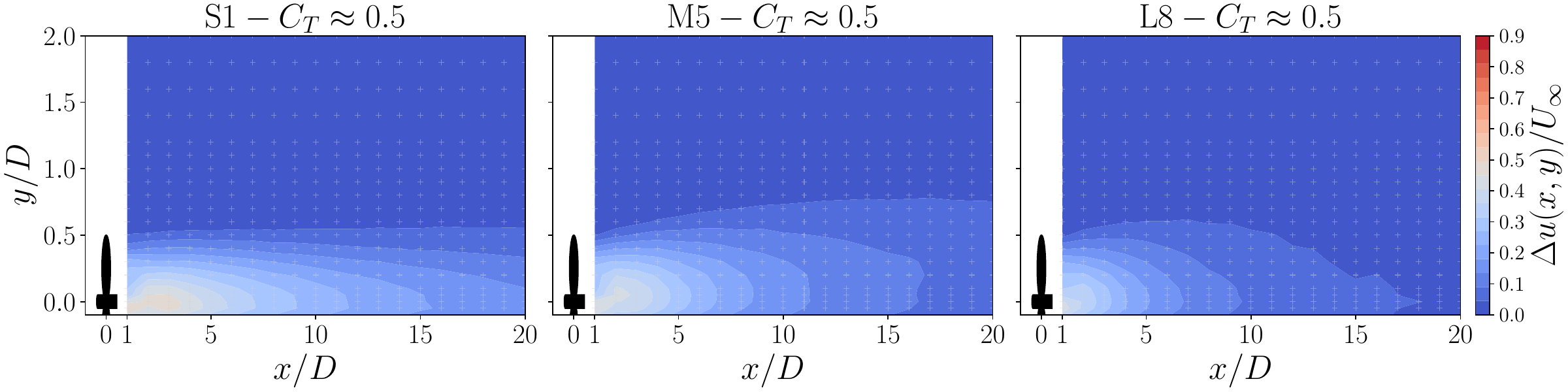}
    \end{subfigure} \\
     \begin{subfigure}[c]{\linewidth}        
        \emph{(b)} \par \vspace{0.1cm}
        \centering
        \includegraphics[width=\linewidth]{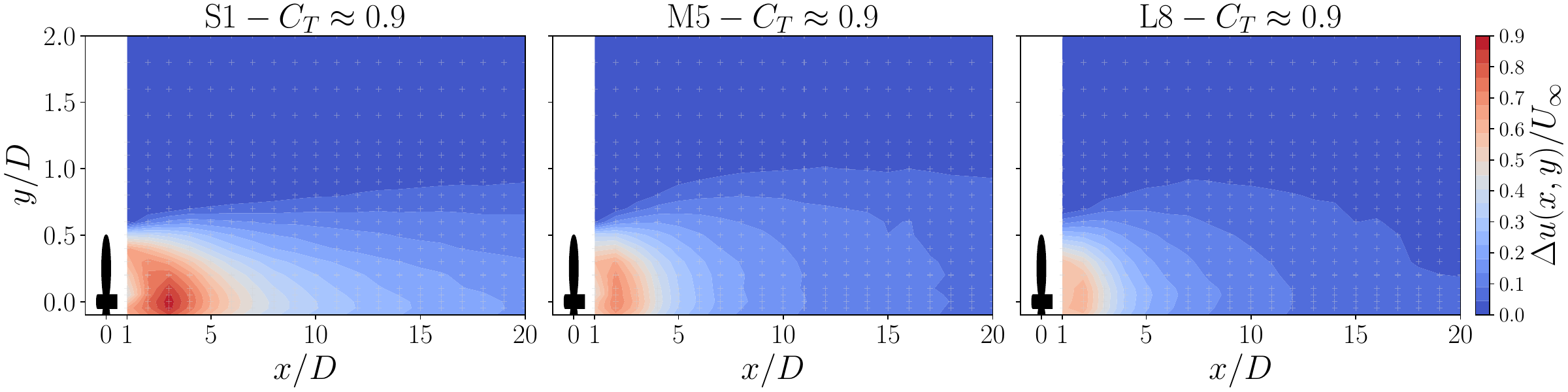}
    \end{subfigure} 
        \caption{Normalised mean streamwise velocity-deficit fields, $\upDelta u/U_{\infty}$. \emph{(a)} $C_T\approx0.5$. \emph{(b)} $C_T\approx0.9$.}
        \label{fig:velo_field}
\end{figure}

\section{Mean-flow and turbulence fields \label{sec:mean_flow_and_turb_fields} }

\begin{figure}
    \begin{subfigure}[c]{\linewidth}
    \emph{(a)} \par \vspace{0.1cm}
            \centering
        \includegraphics[width=\linewidth]{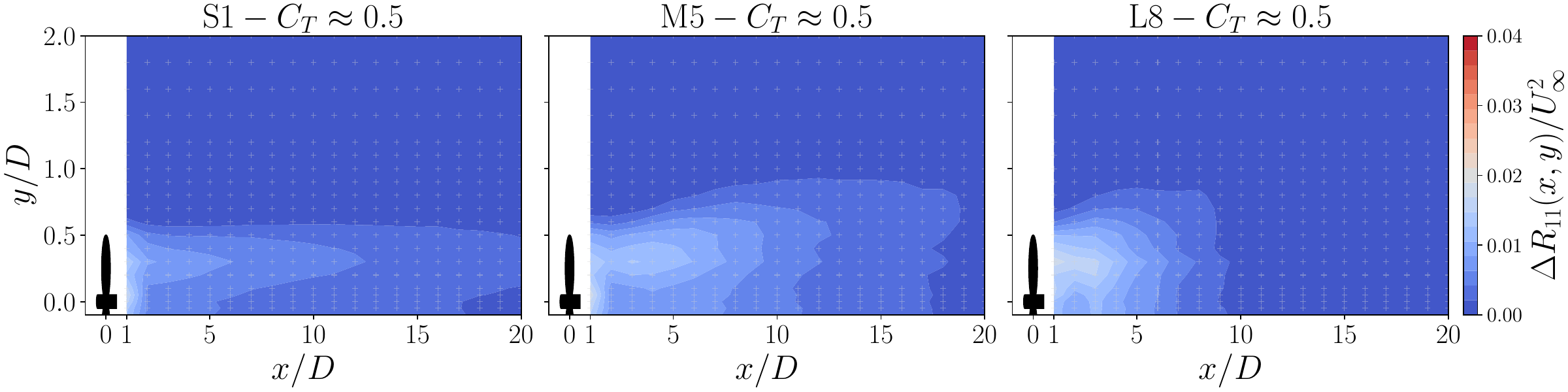}
    \end{subfigure} \\
     \begin{subfigure}[c]{\linewidth}        
        \emph{(b)} \par \vspace{0.1cm}
        \centering
        \includegraphics[width=\linewidth]{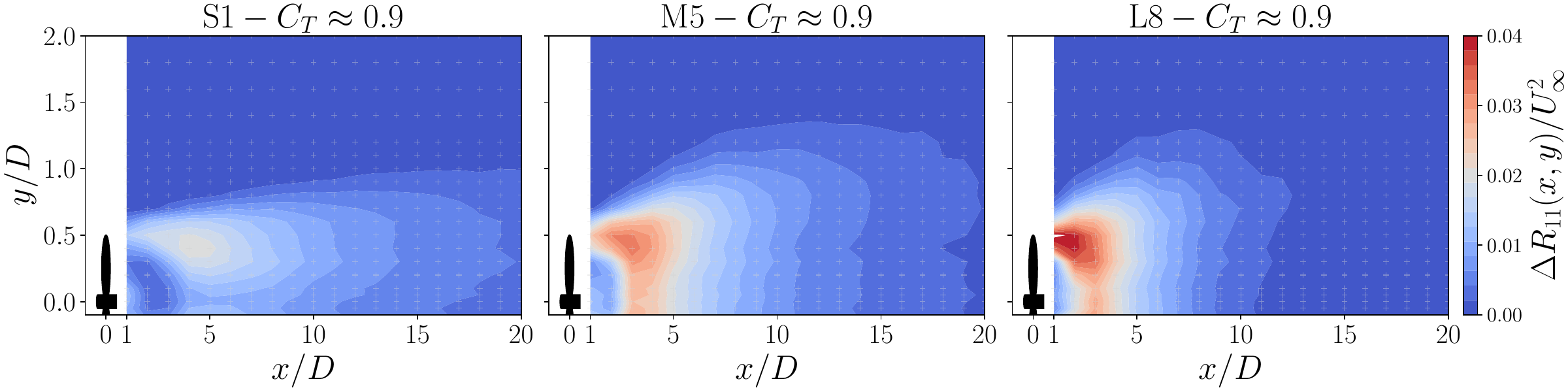}
    \end{subfigure} 
        \caption{Normalised wake-added streamwise Reynolds stress fields, $\upDelta R_{11}/U_{\infty}^2$. \emph{(a)} $C_T\approx0.5$. \emph{(b)} $C_T\approx0.9$. }
        \label{fig:rxx_field}
\end{figure}

\begin{figure}
    \begin{subfigure}[c]{\linewidth}
    \emph{(a)} \par \vspace{0.1cm}
            \centering
        \includegraphics[width=\linewidth]{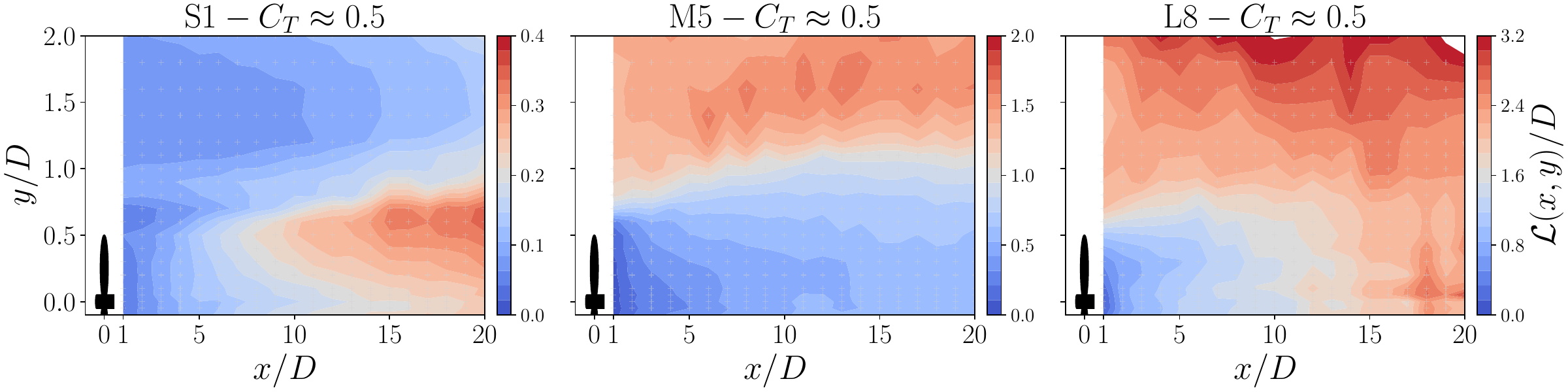}
    \end{subfigure} \\
     \begin{subfigure}[c]{\linewidth}        
        \emph{(b)} \par \vspace{0.1cm}
        \centering
        \includegraphics[width=\linewidth]{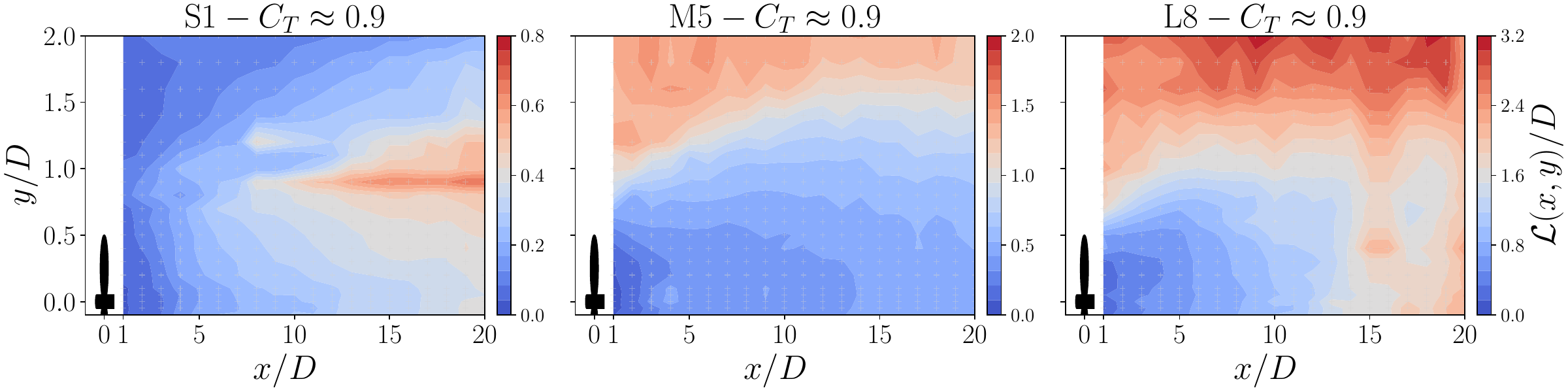}
    \end{subfigure} 
        \caption{Normalised integral length scale fields, ${\cal L}/D$. \emph{(a)} $C_T\approx0.5$. \emph{(b)} $C_T\approx0.9$. Note the change in the ordinate axis between free-stream turbulence cases.}
        \label{fig:ils_field}
\end{figure}

We begin by appreciating qualitatively the influence of free-stream turbulence and the thrust coefficient on the wake structure and its turbulence characteristics. Figures~\ref{fig:velo_field}, \ref{fig:rxx_field}, and \ref{fig:ils_field} present the time-averaged velocity deficit fields $\upDelta u(x,y)$, the wake-added streamwise normal Reynolds stress fields $\upDelta R_{11}(x,y)$, and the streamwise turbulence integral length scale fields ${\cal L}(x,y)$, for three inflow conditions (S1, M5, and L8), with the turbine operating at low and high thrust coefficients ($C_T = 0.5$ and $0.9$). The mean velocity deficit is defined as $\upDelta u(x,y) = U_{\infty}(x) - U(x,y)$, where the local free-stream velocity $U_{\infty}(x)=\langle U(x,y) \rangle_{y/D \in \{1.8,2\}}$ accounts for the small variations of the background flow velocity along the wind-tunnel test section. The wake-added streamwise normal Reynolds stress is defined as $\upDelta R_{11} (x,y) = R_{11} (x,y) - R_{11,\infty}(x)$, where $R_{11}(x,y) = \langle u(x,y)^{'2}\rangle$, and $R_{11,\infty}(x) = \textrm{min}_y (\langle u(x,y)^{'2}\rangle)$ denotes the background turbulence level. This wake-added formulation is commonly used in wind-turbine wake modelling \citep[e.g.][]{Bastankhah2024,Blondel2026}. The streamwise turbulence integral length scale in the wake, ${\cal L}(x,y)$, is estimated by following the same procedure as for the incoming flow (see \S~\ref{sec:Experimental setup}), by integrating the normalised temporal autocorrelation function and using Taylor's hypothesis with the local mean flow velocity. Taylor’s hypothesis should, however, be used with caution in the near field when estimating the size of large-scale vortices, as the turbulence intensity is high.

The presence of FST strongly affects wake development, modifying the wake structure from the near field onward. Increasing $I_{\infty}$ accelerates both the onset and the rate of recovery of the mean velocity deficit and promotes more rapid near-field wake expansion. For low turbulence-intensity inflows (e.g. S1) the velocity deficit remains larger beyond the pressure-recovery region ($x/D \gtrsim 3$), and the wake expands continuously with streamwise distance. As the FST intensity increases, the velocity deficit decays more rapidly, recovery initiates closer to the turbine, and the outer deficit contour contracts in the far field (e.g. for $x/D\gtrsim8$ in case L8). As expected, increasing the thrust coefficient $C_T$ intensifies the initial velocity deficit and enhances the near-field wake width and growth rate. The inflow turbulence integral length scale, although secondary to $I_{\infty}$, also influences wake recovery. For comparable turbulence intensity, larger ${\cal L}_{\infty}$ leads to a delayed onset and slower recovery (e.g. from the comparison of cases S4, M5 and L6) \citep{Bourhis2025}.

The wake-added streamwise normal stress field exhibits a region of enhanced $\upDelta R_{11}(x,y)$ within the annular high-shear tip region ($y/D \approx 0.4-0.6$). The magnitude of $\upDelta R_{11}(x,y)$ in the near field increases with both $C_T$ and $I_{\infty}$, reflecting the strengthening of the tip-shear layer. For sufficiently high turbine loading ($C_T\geq0.7$), an initial increase in $\upDelta R_{11}$ is observed before its downstream decay, likely reflecting the inward spreading of turbulence from the stronger tip-shear layer towards the wake centreline. Increasing $I_{\infty}$ accelerates this near-field turbulence development, bringing the region of high $\upDelta R_{11}$ closer to the turbine, whereas larger ${\cal L}_{\infty}$ shifts this region further downstream. Qualitatively, the outer $\upDelta R_{11}$ contours spread into the ambient flow more rapidly than the outer mean velocity-deficit contour, particularly in the presence of background turbulence, suggesting comparatively more efficient near-field entrainment of turbulent kinetic energy than momentum \citep{Buxton_Chen_2023}. Finally, increasing $I_{\infty}$ significantly accelerates the dissipation of wake-added turbulence such that $\upDelta R_{11} \approx0$ within a few diameters for the highest-$I_{\infty}$ cases (e.g. $\upDelta R_{11}/U_{\infty}^2 \leq0.0015$ for $x/D>10$ in $\{\textrm{L8}; C_T\approx0.5\}$). Hence, whilst an appreciable $\upDelta R_{11}$ persists up to the furthest downstream station under low-turbulence-intensity inflows (e.g. S1), strong background turbulence rapidly suppresses the wake-added contribution, consistent with enhanced turbulent diffusion and more rapid homogenisation with the surrounding flow.

Finally, comparison of the integral length scale fields, ${\cal L}(x,y)$, reveals clear differences in the large-scale wake structures across inflow conditions. For inflows with turbulence integral length scales larger than the turbine diameter (i.e. cases M\# and L\#) the turbine filters the large-scale, energy-containing vortices, resulting in ${\cal L}(x,y) \leq {\cal L}_{\infty}$. As the wake develops downstream, ${\cal L}(x,y)$ gradually increases and approaches the background value (${\cal L} \to {\cal L}_{\infty}$) with this accelerated ``recovery'' commensurate with both $I_{\infty}$ and $\cal L_{\infty}$ increasing. This behaviour contrasts with cases S\#, where ${\cal L}_{\infty}\leq D/2$. In these cases ${\cal L}(x,y)\geq {\cal L}_{\infty}$, and a region of elevated turbulence integral length scale emerges within the tip-shear layer. The magnitude of ${\cal L}(x,y)$ in this region increases with the thrust coefficient and as the wake develops. Interestingly, the onset of this growth of ${\cal L}$ in the tip-shear layer approximately coincides with the streamwise location where high-frequency tip vortices disappear from the turbulence spectra (not shown here for brevity; see \cite{Bourhis2025}). Akin to porous-body tip-shear-layer dynamics \citep{Cicollin2024,Li2024,Bourhis2024}, tip vortices initially confined within the annular tip-shear layer interact and merge, forming progressively larger structures as the shear layer develops, thereby increasing ${\cal L}$. These structures exhibit a characteristic broadband low-frequency peak in the turbulence spectra at a Strouhal number of $St\approx0.2$, corresponding to the meandering of the tip-shear layer \citep{Yang2019_2,Yang2019,Bourhis2025}. Consistently, increasing turbine loading intensifies the tip-shear layer and promotes the development of larger structures within it.

To ensure that the evolution of ${\cal L}_x(x,y)$ is not artificially driven by local mean-velocity recovery, the integral temporal scale ${\cal T}(x,y)={\cal L}(x,y)/U(x,y)$ was also examined and showed qualitatively similar behaviour, including tip-shear-layer peaks in cases S\# and recovery towards ambient conditions for cases M\# and L\#. This concludes the qualitative comparison of the mean-flow and turbulence fields; further quantitative analysis, including the effects of FST and $C_T$ on mean-flow recovery and wake dynamics is presented in \cite{Bourhis2025}.

\section{Self-similar behaviour of wind-turbine wakes \label{sec:self_sim}}

Having qualitatively examined the wake mean-flow and turbulence fields, we now investigate the extent to which wind-turbine wakes embedded in external turbulence exhibit self-similar behaviour, which underpins the equilibrium similarity analysis of \citet{George1989}.

As George’s framework for statistically axisymmetric turbulent wakes assumes statistical axisymmetry of turbulent quantities, wind turbine wake axisymmetry is first considered. Although commonly assumed in wind-turbine wake modelling, wakes are only approximately axisymmetric, particularly in the near field, owing to tower effects, ground proximity and vertical shear in the atmospheric boundary layer \citep{Sanderse2011,Abraham2019}. In the present experiments, these effects were mitigated by elevating the turbine hub to $1$~m above the wind-tunnel floor and measuring in the horizontal plane (deviations from symmetry primarily arise in the vertical direction). In addition, the presence of spatially uniform background turbulence, without vertical velocity shear, promotes wake axisymmetry by weakening the imprint of the tower on the wake \citep{Wu2012}. Finally, the horizontal symmetry of all 24 wakes was confirmed from the mean flow profiles (velocity, turbulence intensity, dissipation rate) using the hot-wire probes located on the negative-y side of the wake. The wakes are therefore assumed to be statistically axisymmetric, and the horizontal coordinate is treated as the radial coordinate, $y\equiv r$. 

\begin{figure}
    \centering
    \includegraphics[width=0.9\linewidth]{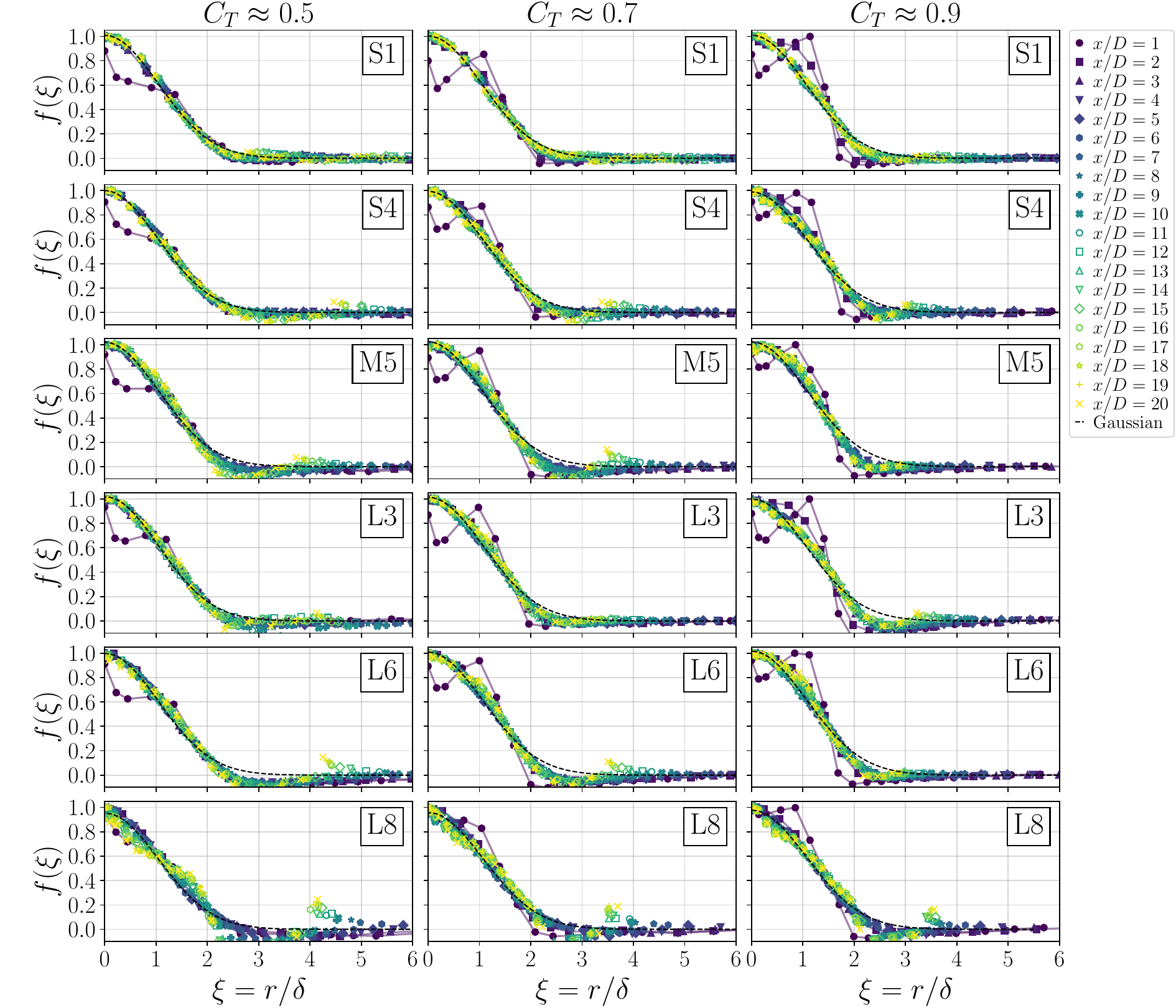}
    \caption{Time-averaged velocity-deficit profiles at different streamwise distances, shown in self-similar coordinates, where $f(\xi) =\upDelta u(x,r)/ \upDelta u_0(x)$.}
    \label{fig:velo_def_simi}
\end{figure}

\begin{figure}
    \centering
    \includegraphics[width=0.9\linewidth]{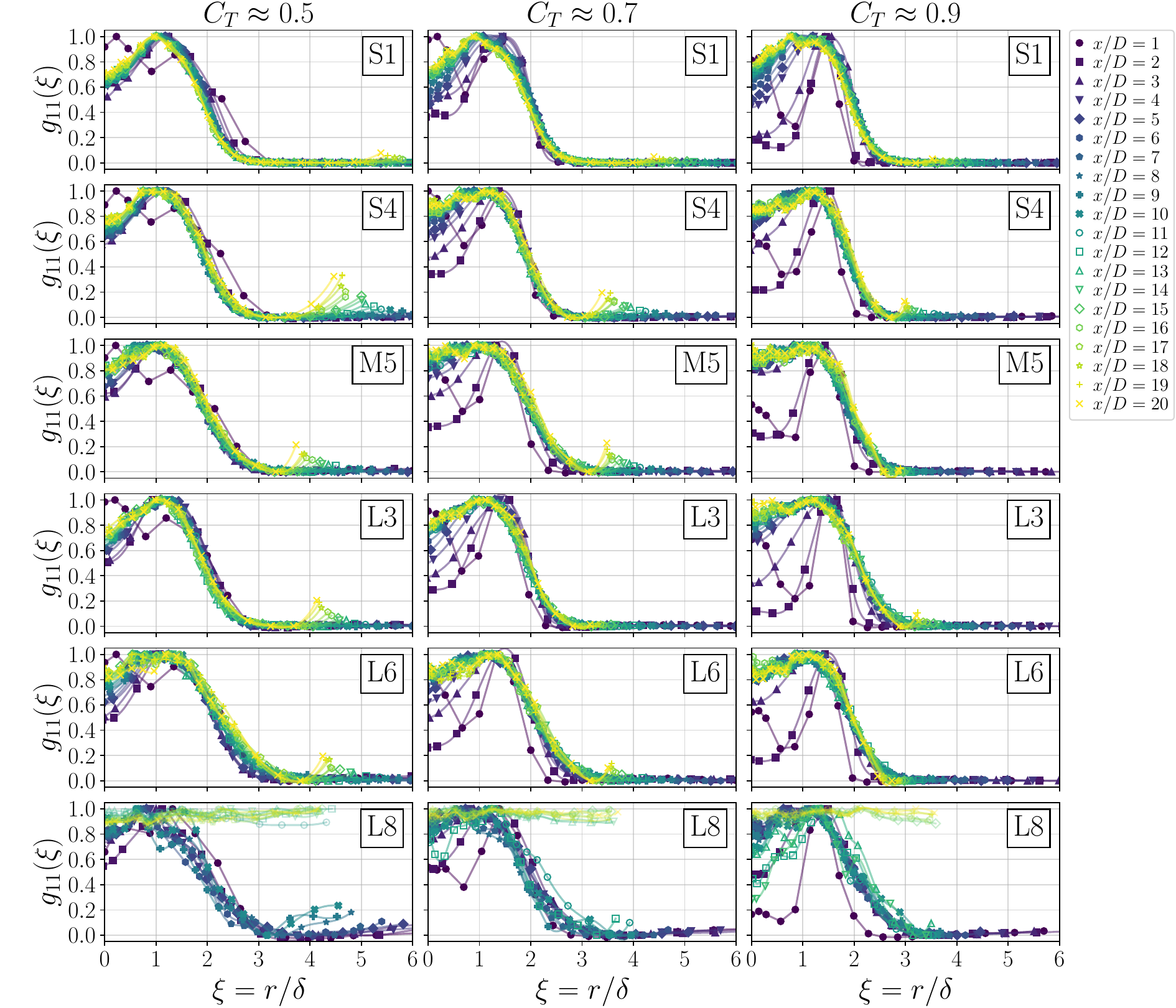}
    \caption{Streamwise normal Reynolds stress profiles at different streamwise distances, shown in self-similar coordinates, where $g_{11}(\xi) = \upDelta R_{11}(x,r)/\max_r(\upDelta R_{11}(x,r))$. For case L8, $g_{11}(\xi)$ is replaced by $g_{11}(\xi) = R_{11}(x,r)/\max_r(R_{11}(x,r))$ when the wake-added contribution falls below $\upDelta R_{11}/U_{\infty}^2 < 0.0015$ (semi-transparent lines).}
    \label{fig:rxx_simi}
\end{figure}

\begin{figure}
    \centering
    \includegraphics[width=0.9\linewidth]{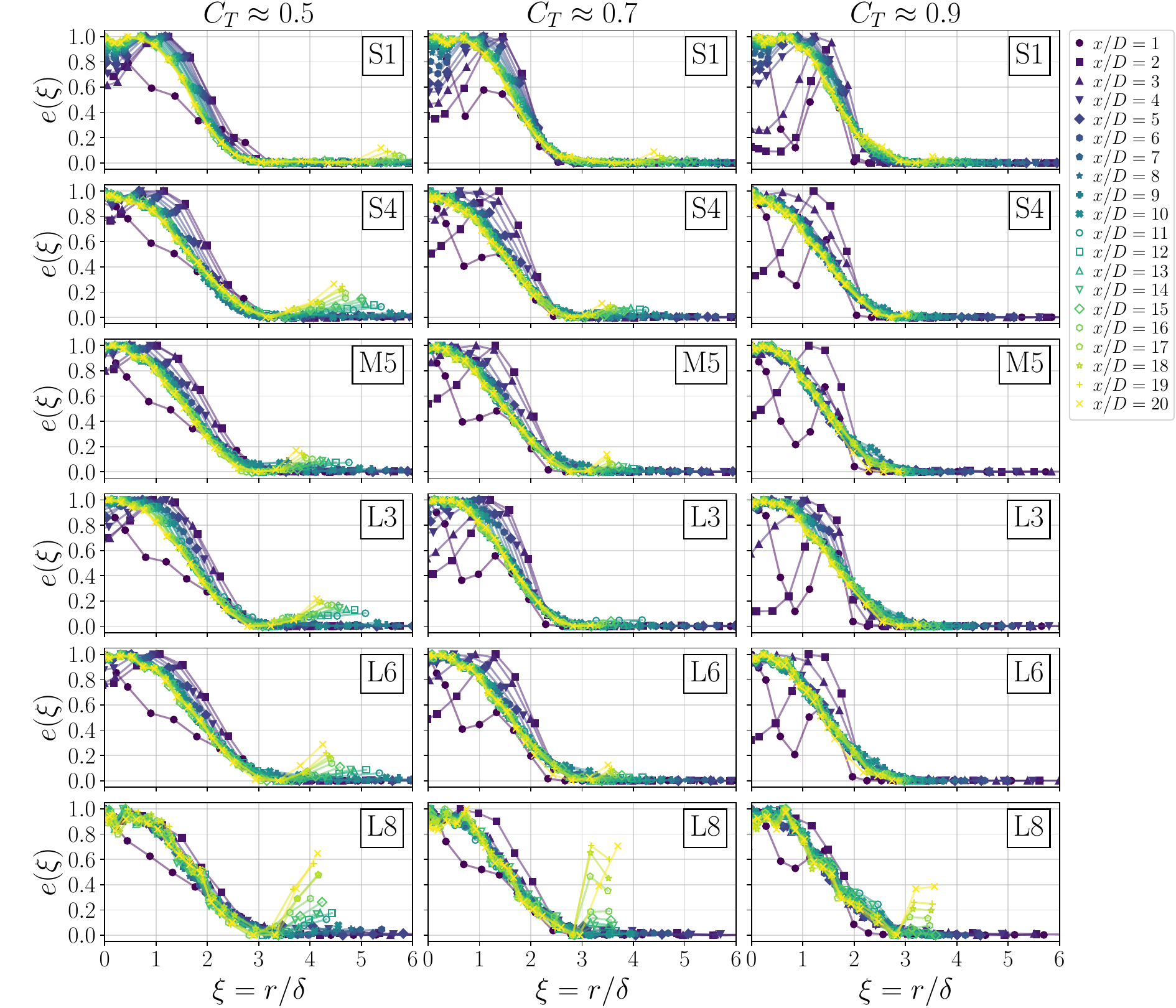}
    \caption{Dissipation rate profiles at different streamwise distances, shown in self-similar coordinates, where $e(\xi) = \upDelta \varepsilon(x,r)/\max_r(\upDelta \varepsilon(x,r))$.}
    \label{fig:eps_simi}
\end{figure}

The self-preservation of turbulent quantities in the 24 wakes is assessed for the mean velocity deficit (figure~\ref{fig:velo_def_simi}), the streamwise normal Reynolds stress (figure~\ref{fig:rxx_simi}), and the dissipation rate (figure~\ref{fig:eps_simi}). We begin with the mean velocity deficit, for which radial profiles are normalised by the centreline velocity deficit, approximated by the maximum velocity deficit, $\upDelta u_0(x) = \max_r(\upDelta u(x,r))$, and plotted as functions of the scaled radial coordinate $\xi = r/\delta(x)$. Here, $\delta(x)$ is the integral wake width obtained from the mean profiles as $\delta^2(x) = 1/\upDelta u_0(x) \int_0^{r_0} \upDelta u(x,r) r dr$, where $r_0$ is the radial location at which $\upDelta u(x,r) = 0.05 \upDelta u_0(x)$ (other thresholds were tested and yield similar trends). The resulting self-similar profiles $f(\xi) = \upDelta u(x,r)/ \upDelta u_0(x)$ for six turbulent background flows are shown in figure~\ref{fig:velo_def_simi}. 

The mean velocity-deficit profiles reach self-similarity close to the turbine, becoming self-similar for $x/D \geq 3$ under all turbulent inflow conditions, and at even shorter downstream distances at low thrust coefficient. Consistently, \cite{Xie2015} reported a rapid onset of horizontal self-similarity of $f(\xi)$ in their LES of wind-turbine wakes exposed to neutral ABL turbulence with vertical shear (for $x/D \geq 2$). FST further promotes this early transition by enhancing turbulent mixing and accelerating the breakdown of entrainment-inhibiting tip vortices \citep{Lignarolo2015,Bourhis2025,Couliou2026}. Whilst self-similarity is preserved up to $x/D=20$ for low to moderate turbulence intensity inflows ($I_{\infty}\leq5.8\%$), the two high-turbulence intensity cases (L7 \& L8) exhibit a weaker profile collapse in the far-field. In these cases, the residual velocity deficit becomes smaller than the background flow velocity fluctuations ($\upDelta u_0 < \sqrt{\langle u_{\infty}^{'2}} \rangle$); for example, $\upDelta u_0 (x/D=20) \approx 0.4 \sqrt{\langle u_{\infty}^{'2}} \rangle$ for case L8 at $C_T=0.5$. By contrast, for low-turbulence inflows the velocity deficit remains much larger than the ambient fluctuations ($\upDelta u_0 \gg \sqrt{\langle u_{\infty}^{'2}})\rangle$; for example, $\upDelta u_0 (x/D=20) \approx 20 \sqrt{\langle u_{\infty}^{'2}} \rangle$ for case S1 at $C_T=0.5$. Hence, under high-turbulence conditions, the far-wakes become ``weak'' relative to the background turbulence, potentially hindering a strictly self-similar evolution of the mean velocity deficit \citep{Johnson2014}. A Gaussian fit to the self-similar profiles is shown for reference, as most recent wind turbine wake models assume a Gaussian mean velocity deficit \citep[e.g.][]{Bastankhah2014,Bastankhah2024}. The largest deviations occur near the wake edge, within the tip-shear layer, where strong velocity gradients and intermittency affect local mixing and entrainment processes that are not represented by the smooth, diffusion-based structure implicit in a Gaussian profile \citep{Neunaber2020,Zheng2023}.

Self-similar forms for the streamwise normal Reynolds stress and the dissipation-rate profiles are next examined. Dissipation rates in the wakes, $\varepsilon(x,r)$, are estimated following the procedure described in \S~\ref{sec:Experimental setup}, by integrating the dissipation spectra after modelling the high-wavenumber range. For wakes developing in a non-turbulent background, self-similarity of these quantities is classically assessed using normalised profiles of the form $g_{11}(\xi) = R_{11}(x,r)/\max_r(R_{11}(x,r))$ and $e(\xi) =\varepsilon(x,r)/\max_r(\varepsilon(x,r))$ \citep{George1989,Johansson2003,Dairay2015}. This conventional formulation, however, entails a slight inconsistency in the similarity scaling, since the mean velocity deficit is normalised by its centreline value, whereas $R_{11}$ and $\varepsilon$ are scaled by maxima located within the tip-shear layer. A further complication arises in the presence of FST, as $R_{11}$ and $\varepsilon$ progressively adjust to background levels as the wake mixes with the free stream, causing $g_{11}(\xi)\to 1 $ and $e_{11}(\xi) \to 1$. To assess self-similarity in the presence of external turbulence, we therefore adopt an alternative formulation based on quantities rescaled relative to the free-stream levels. Similarly to the wake-added streamwise normal Reynolds stress ($\upDelta R_{11} (x,r)$), the turbulence dissipation rate is rescaled relative to the background level as $\upDelta \varepsilon(x,r) = \varepsilon(x,r) - \varepsilon_{\infty}(x)$, where $\varepsilon_{\infty}(x) = \textrm{min}_r (\varepsilon(x,r))$. Self-similarity is then assessed using the normalised forms $g_{11}(\xi) = \upDelta R_{11}(x,r)/\max_r(\upDelta  R_{11}(x,r))$, and $e(\xi) = \upDelta \varepsilon(x,r)/\max_r(\upDelta \varepsilon(x,r))$, shown in figures~\ref{fig:rxx_simi} and \ref{fig:eps_simi} for six turbulent inflows.

For low to moderate FST intensity inflows (S1--L6), the profiles of $g_{11}(\xi)$ collapse onto a single well-defined saddle-shaped structure, characteristic of self-similar behaviour \cite[e.g.][]{Pope2000}, although this regime is attained further downstream than for the mean velocity deficit. Increasing FST intensity shortens the streamwise distance required to reach self-similarity, while no systematic dependence on the FST integral length scale is observed. For the highest turbulence intensity cases (L7 \& L8), self-similarity of $g_{11}(\xi)$ is less clearly established. Although a saddle-shaped structure remains discernible in the near field, the increased scatter of the collapsed profiles indicates a weaker self-similar behaviour. Compared with lower-turbulence-intensity cases, self-similarity persists over a much shorter streamwise extent, primarily due to the rapid dissipation of turbine-induced turbulence. In the far field, the wake-induced streamwise normal Reynolds stress becomes indistinguishable from the background within the hot-wire uncertainty; hence, for $\upDelta R_{11} \leq 0.0015$, the semi-transparent profiles shown correspond to the classical self-similar form $R_{11}(x)/\max_r(R_{11}(x,r))$ and illustrate convergence towards background levels ($\upDelta R_{11}\approx0$). Therefore, while self-similarity of $f(\xi)$ is maintained over the entire measurement domain, self-similarity of $g_{11}(\xi)$ breaks down more rapidly, reflecting the faster dissipation of turbine-induced turbulence kinetic energy relative to the recovery of the velocity deficit, a behaviour also reported in the LES of wind turbine wakes exposed to external turbulence by \citet{Zhang2023}. These observations are also closely related to the experimental study of \citet{Rind2012} on axisymmetric solid disk wakes exposed to FST, where sufficiently strong background turbulence ($I_{\infty} \geq 3.3\%$) was shown to suppress self-similarity of the turbulence normal stresses in the far field ($x/D\geq65$). However, while \citet{Rind2012} focused exclusively on the far wake---where wake-added turbulence had largely decayed---the present measurements extend into the near and intermediate wake, where wake turbulence remains strong relative to ambient turbulence. In this region, we found that self-similarity of $g_{11}(\xi)$ is well preserved at moderate and high FST intensity, however, at high turbulence intensity it breaks down rapidly downstream, with the normal stresses gradually becoming radially uniform and reaching the free-stream level.

Figure~\ref{fig:eps_simi} shows the self-similar form of the rescaled turbulence dissipation rate, $e(\xi)$. For all inflow conditions, the dissipation profiles transition from an initial development region to a self-similar regime. As observed for $f(\xi)$ and $g_{11}(\xi)$, increasing FST intensity accelerates this transition, while the free-stream integral length scale plays a secondary role. Once self-similarity is established, the radial profiles collapse more clearly for low to moderate FST intensities (S1--L6) than for the most turbulent inflows (L7 \& L8). Nevertheless, an approximately self-similar shape is maintained up to the last measurement station even for the highest turbulence intensities, in contrast to $g_{11}$. It can be noted that the elevated $e(\xi)$ values in the outer wake for case L8 likely reflect the proximity of the corresponding hot wires to the tunnel walls, locally degrading the self-similar collapse. The weaker collapse observed at high turbulence intensities is also associated with the reduced magnitude of wake-added dissipation in the far field : as $\varepsilon \to \varepsilon_{\infty}$, the normalised dissipation $e(\xi)$ becomes increasingly sensitive to small relative variations, leading to a degradation of the profile collapse (e.g. at $x/D=20$ and $C_T=0.5$, $\varepsilon \approx 1.3\varepsilon_{\infty}$ for L8, whereas $\varepsilon\approx 70\varepsilon_{\infty}$ for S1). It further emphasises the progressive dominance of ambient turbulence over wake-induced turbulence and the rapid vanishing of the wake in the presence of strong ambient turbulence.

\section{Scaling of the time-averaged turbulent energy dissipation rate \label{sec:Scaling of the time-averaged turbulent energy dissipation rate}}

We now examine the scaling of the energy dissipation rate across the 24 wakes to assess how external turbulence and initial flow disturbances (i.e. mean-flow momentum deficit, turbine-added turbulence and coherent dynamics, fixed by $C_T$ and the TSR) modify the turbulence nature within the wake (equilibrium vs. non-equilibrium turbulence dissipation scaling). The dissipation coefficient, $C_{\varepsilon}$, is computed without using K41 similarity or scaling assumptions, as $C_{\varepsilon} = \varepsilon {\cal L}/{\langle u^{\prime 2} \rangle}^{3/2}$ (formulation used by \citet{Valente2012,Goto2016b,Mora2019}). As, in our case, two streams of turbulence coexist and interact---the wake and the ambient turbulence---we will examine the state of the turbulence in two different regions : the wake core, near the centreline, where turbine-generated turbulence contributes \emph{a priori} predominantly to the local turbulence; and the outer wake, where several flow phenomena may modify the dissipation scaling, notably (i) the proximity of the ambient turbulent environment, which locally alters turbulent–turbulent interface dynamics and entrainment processes \citep{Chen_Buxton_2024,Alves2025}; (ii) turbine-induced coherent structures, such as tip vortices and structures arising from vortex interactions \citep{Bourhis2025}; and (iii) intermittency, which may further affect dissipation behaviour \citep{Schmitt2025}.

\begin{figure}
\centering
    \begin{subfigure}[c]{0.94\linewidth}
        \emph{(a)} \par \vspace{0.1cm}
        \centering
        \includegraphics[width=\linewidth]{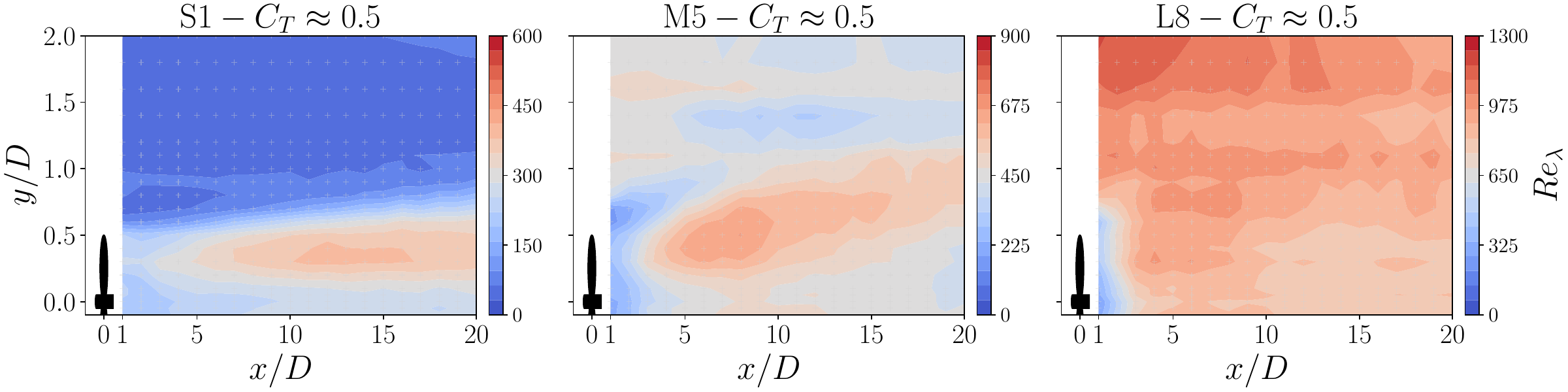}
    \end{subfigure}
    \\
    \begin{subfigure}[c]{0.94\linewidth}        
        \emph{(b)} \par \vspace{0.1cm}
        \centering
        \includegraphics[width=\linewidth]{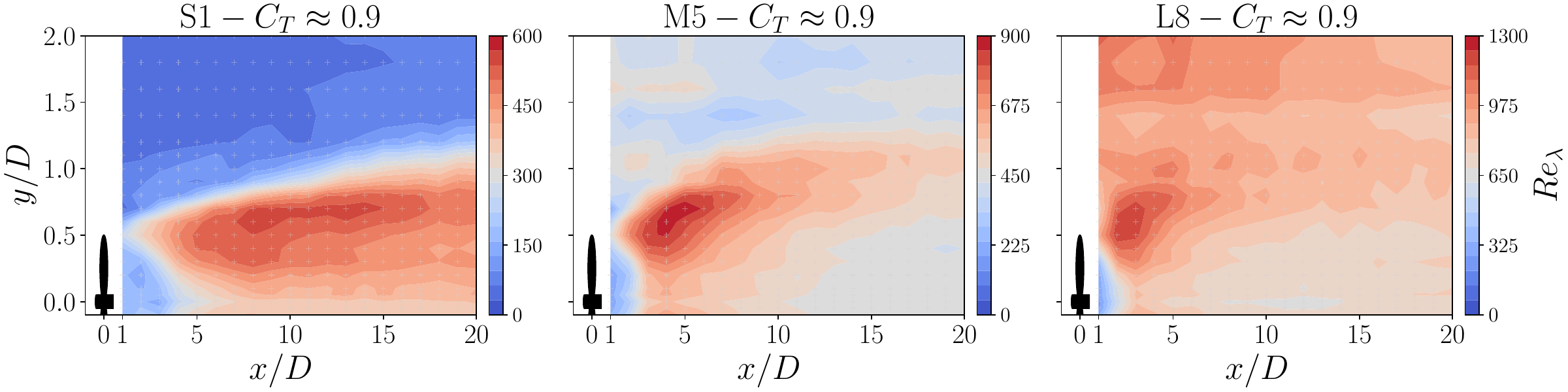}
    \end{subfigure}
    \caption{Taylor-scale Reynolds number fields, $Re_{\lambda}$. \emph{(a)} $C_T\approx0.5$. \emph{(b)} $C_T\approx0.9$. Note the change in the ordinate axis between free-stream turbulence cases. }
    \label{fig:Re_lambda_field}
\end{figure}

\begin{figure}
\centering
    \begin{subfigure}[c]{0.94\linewidth}
        \emph{(a)} \par \vspace{0.1cm}
        \centering
        \includegraphics[width=\linewidth]{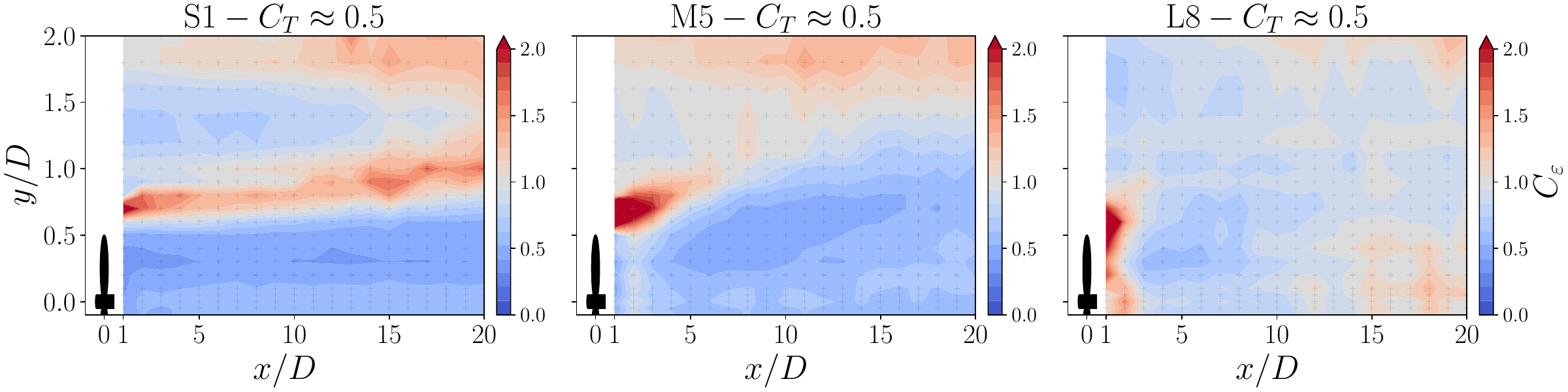}
    \end{subfigure}
    \\
    \begin{subfigure}[c]{0.94\linewidth}        
        \emph{(b)} \par \vspace{0.1cm}
        \centering
        \includegraphics[width=\linewidth]{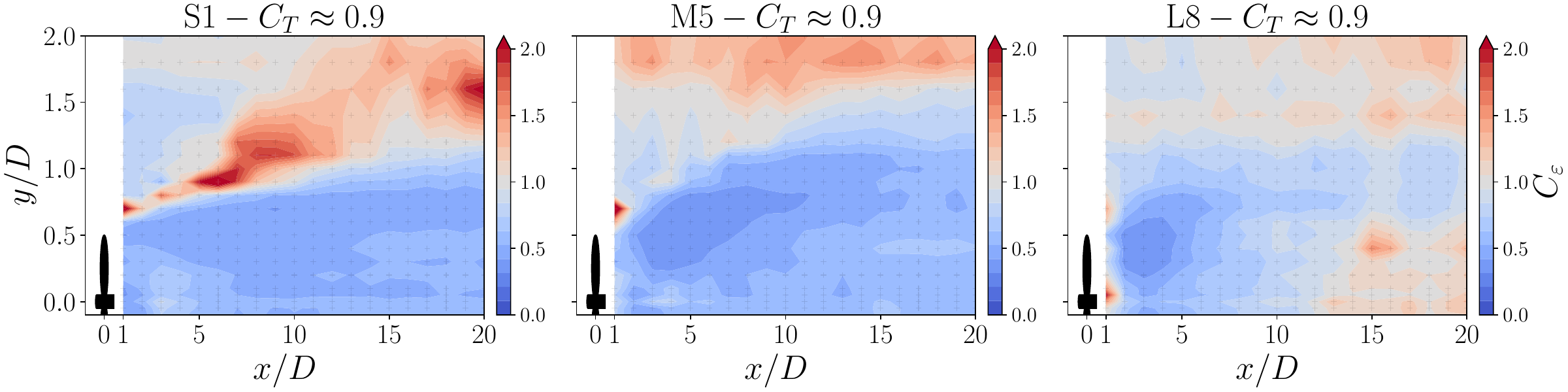}
    \end{subfigure}
    \caption{Normalised dissipation coefficient fields, $C_{\varepsilon}$. \emph{(a)} $C_T\approx0.5$. \emph{(b)} $C_T\approx0.9$.}
    \label{fig:Ceps_field}
\end{figure}

\begin{figure}
    \centering
    \includegraphics[width=0.94\textwidth]{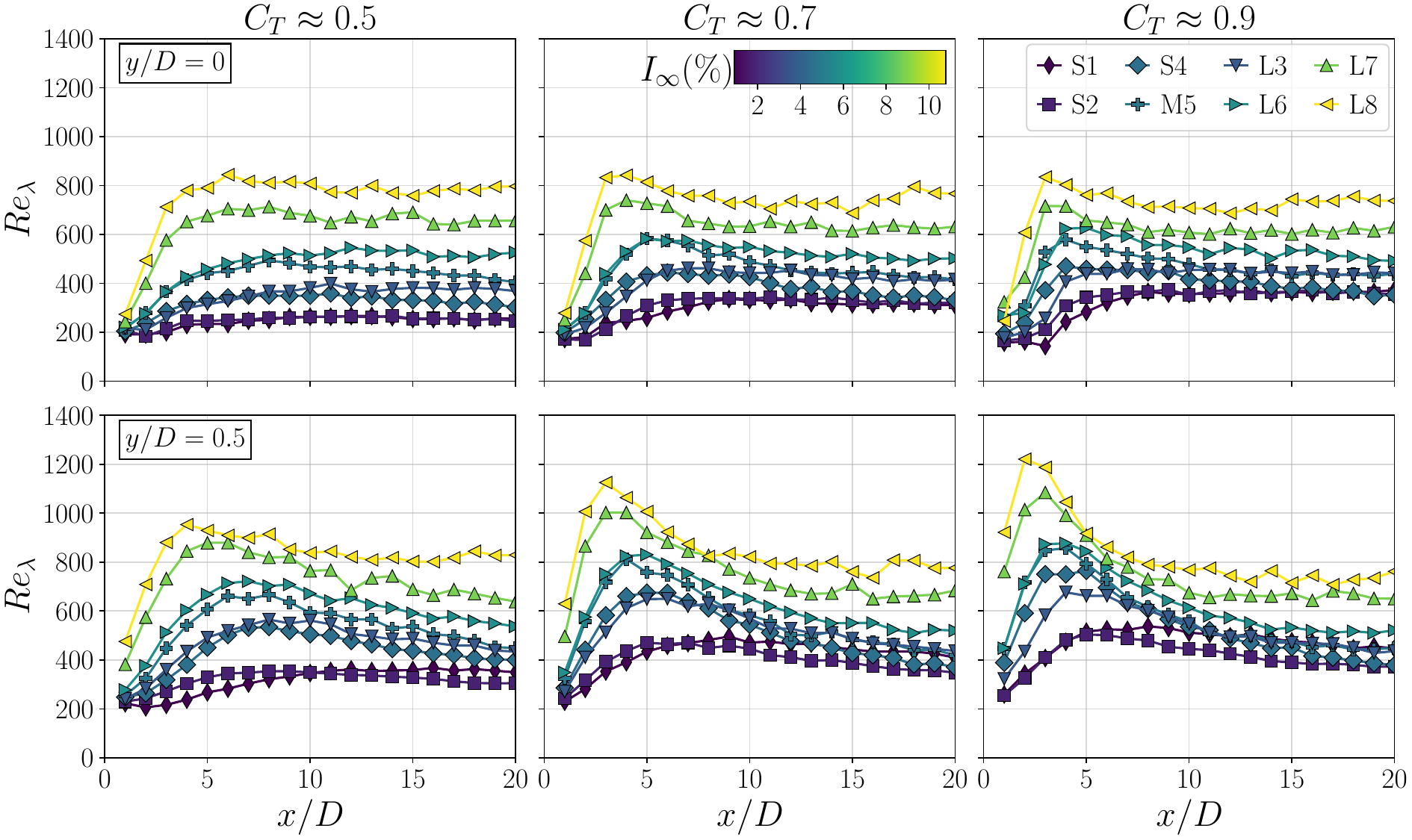}
    \caption{Streamwise evolution of the Taylor-scale Reynolds number, $Re_{\lambda}(x)$, along the wake centreline ($y/D=0$, up) and in the blade-tip region ($y/D=0.5$, bottom).}
    \label{fig:Reynolds_Taylor}
\end{figure}

\begin{figure}
\centering
    \begin{subfigure}[c]{0.94\linewidth}
        \emph{(a)} \par \vspace{0.1cm}
        \centering
        \includegraphics[width=\linewidth]{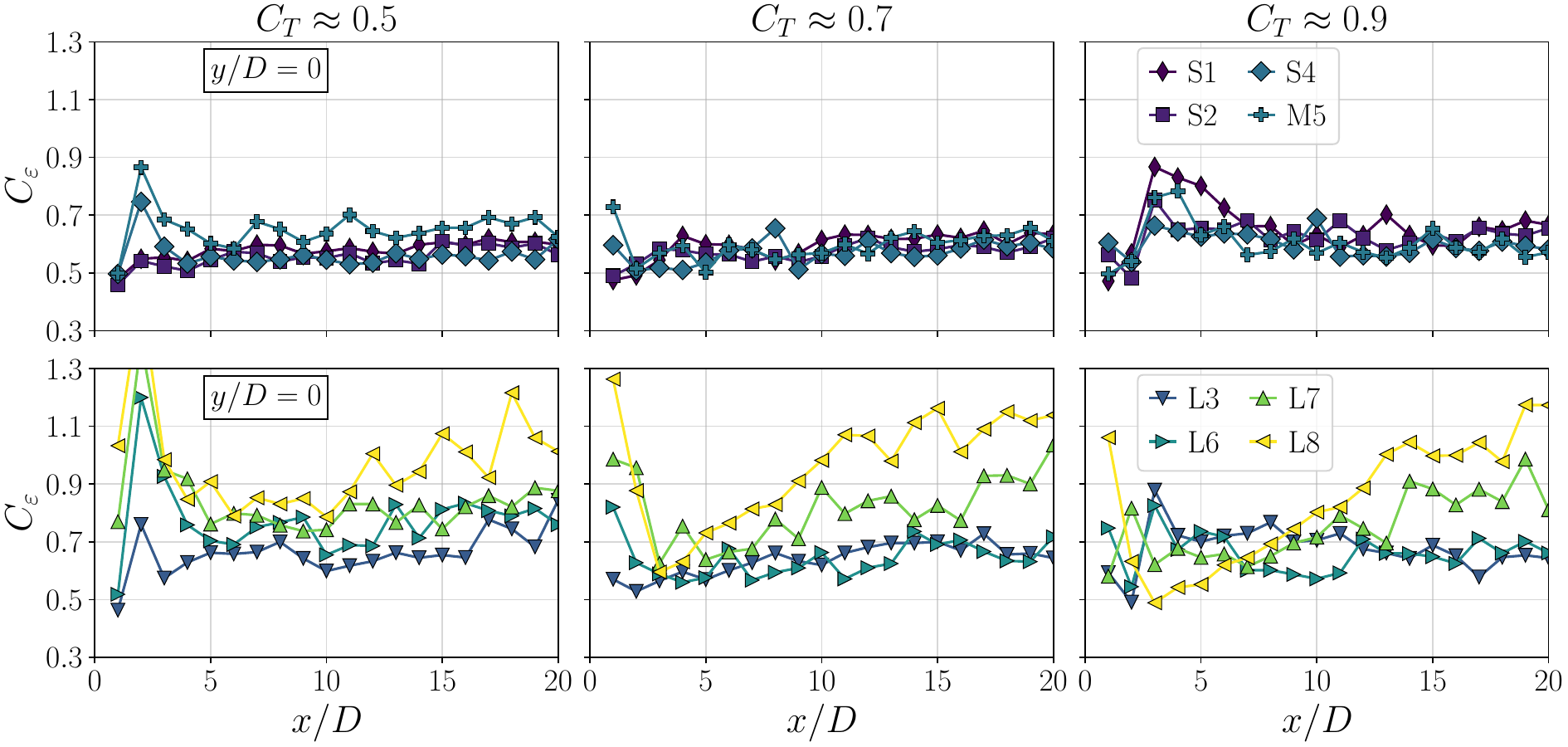}
    \end{subfigure}
    \\
    \begin{subfigure}[c]{0.94\linewidth}        
        \emph{(b)} \par \vspace{0.1cm}
        \centering
        \includegraphics[width=\linewidth]{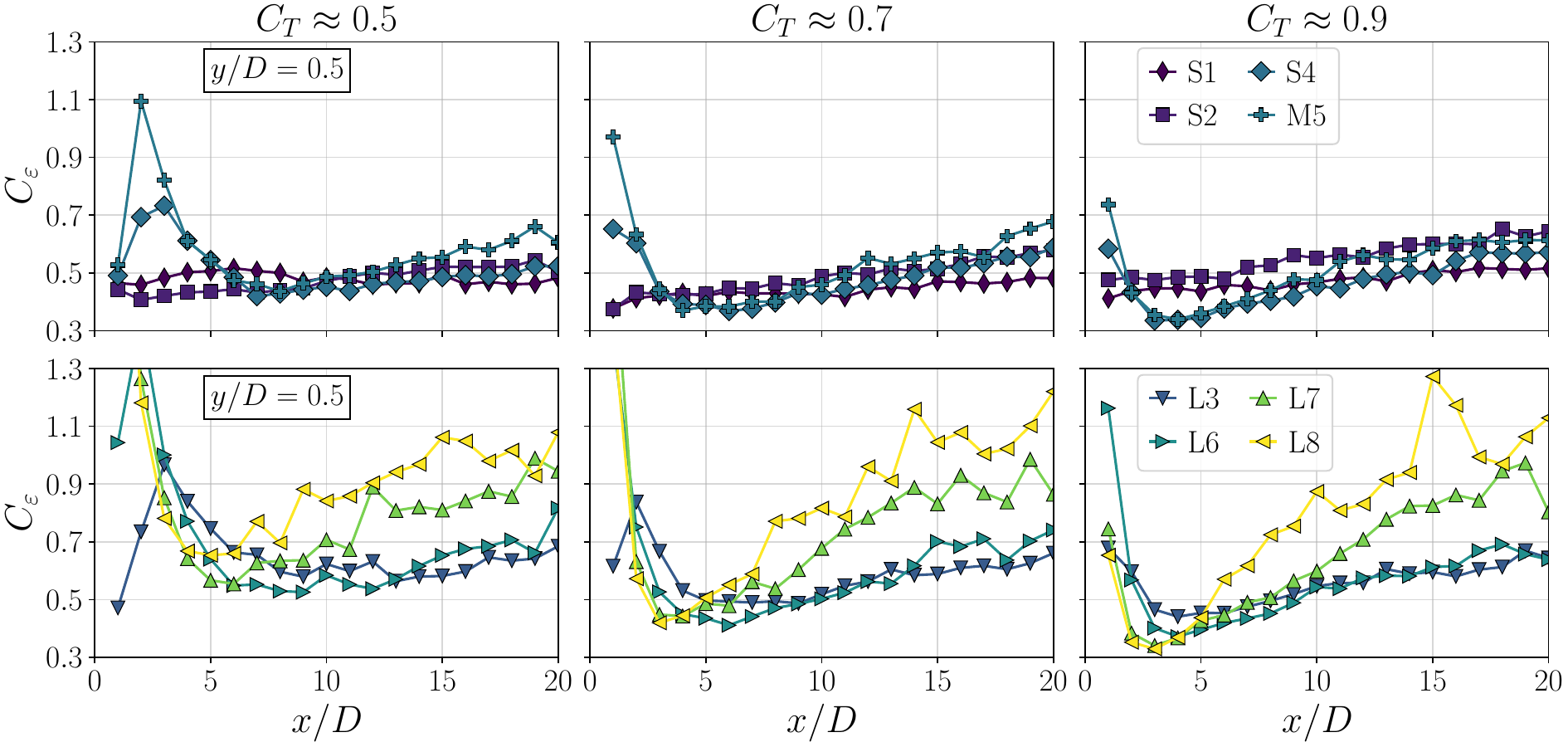}
    \end{subfigure}
    \caption{Streamwise evolution of the normalised dissipation coefficient, $C_{\varepsilon}(x)$, along the wake centreline ($y/D=0$) \emph{(a)} and in the blade-tip region ($y/D=0.5$) \emph{(b)}. FST cases are split into two separate panels for visualisation purposes.}
    \label{fig:Ceps_stream}
\end{figure}

Figures~\ref{fig:Re_lambda_field} and \ref{fig:Ceps_field} present the spatial distributions of the Taylor-scale Reynolds number, $Re_{\lambda}(x,y)$, and the dissipation coefficient, $C_{\varepsilon}(x,y)$, for three FST cases and two thrust coefficients. These fields are complemented by the streamwise evolutions, $Re_{\lambda}(x)$ and $C_{\varepsilon}(x)$, extracted along the centreline ($y/D=0$) and in blade-tip region ($y/D=0.5$) for all inflow conditions and thrust coefficients (figures~\ref{fig:Reynolds_Taylor} and \ref{fig:Ceps_stream}).

The first observation is that the Taylor Reynolds number in the wake is consistently high, exceeding $Re_{\lambda}=200$ everywhere and reaching $Re_{\lambda}\sim O(10^3)$ for the highest FST intensity case (L8). In the wake core, it varies by approximately a factor of three between the lowest- and highest-$I_{\infty}$ cases, highlighting the strong influence of the background turbulence on the inner-wake turbulence. In the low-turbulence cases, $Re_{\lambda}$ in the wake substantially exceeds that of the ambient flow, with a maximum located in the rotor tip-shear layer (e.g. $Re_{\lambda}\approx560$ at $\{x/D,y/D\}=\{8,0.6\}$ for $\{\textrm{S1};C_T=0.9\}$). As the wake spreads downstream, the region of elevated $Re_{\lambda}$ expands outward, highlighting the progressive redistribution of turbine-generated turbulence into the surrounding flow. As FST intensity is increased, $Re_{\lambda}$ rises throughout the measurement domain and the contrast between wake and ambient turbulence weakens, especially at low $C_T$. For the inflow with the highest $I_{\infty}$ (L8), the wake is embedded in an already highly turbulent environment : for $C_T\approx0.5$, $Re_{\lambda}$ in the ambient flow exceeds that within the wake, whereas for $C_T\approx0.9$ a local maximum is visible near the rotor within the tip-shear layer, but $Re_{\lambda,\infty} \geq Re_{\lambda} $ over most of the domain. More generally, for high-turbulence cases (L7 \& L8), the external turbulence rapidly becomes dominant, thereby reducing the relative contribution of internally (turbine-) generated turbulence. 

This trend is also reflected in the streamwise evolution : $Re_{\lambda}$ increases in the near wake, with a faster growth rate at higher $I_{\infty}$. The centreline and tip-region data differ most markedly for $x/D \lesssim 5$, where the coherent motions associated with the tip-vortex system are the most energetic \citep{Bourhis2025,Biswas_Buxton_2026}. Notably, the tip-shear layer systematically exhibits higher $Re_{\lambda}$ than the centreline, particularly in the near field, highlighting the strong contribution of shear-layer turbulence production. In both the tip-shear layer and along the wake centreline, the location of the maximum $Re_{\lambda}$ shifts closer to the turbine as either $C_T$ or $I_{\infty}$ increases. After this initial growth phase, the streamwise evolution of $Re_{\lambda}$ differs between the centreline and the blade-tip region. For most cases, there exists a streamwise location beyond which $Re_{\lambda}(x,y=0)$ remains approximately constant downstream. However, at the wake edge, the initial increase in $Re_{\lambda}$ is followed by a more pronounced and sustained decay. In the far-field ($x/D\gtrsim 10$), where the influence of turbine-generated turbulence weakens, the magnitude of $Re_{\lambda}$ at the wake centreline follows a strict hierarchical ordering with increasing $I_{\infty}$. A similar behaviour is generally observed in the outer wake ($y/D=0.5$); however $Re_{\lambda}(\textrm{S1}) \geq Re_{\lambda}(\textrm{S2})$ for $x/D\gtrsim 10$ at $C_T=0.5$, and $Re_{\lambda}(\textrm{S1}) \geq Re_{\lambda}(\textrm{S2, S4, L3, M5})$ for $C_T=0.9$ in the far field, despite $Re_{\lambda,\infty}(\textrm{S1})$ being lower than in the other cases. This behaviour is likely the result of persistent tip-vortex activity within the tip-shear layer under near-zero FST intensity (S1). The longer survival of tip vortices promotes their merging into larger coherent structures, enriching tip-shear-layer dynamics \citep{Bourhis2025} and increasing the local integral length scale (figure~\ref{fig:ils_field}), thereby limiting the downstream decay of $Re_{\lambda}$. Increasing FST intensity instead accelerates tip-vortex dissipation, inhibiting large-scale structure formation within the tip-shear layer. Consequently, higher ambient turbulence levels do not necessarily translate into larger $Re_{\lambda}$ within the tip-shear layer, because of reduced tip-shear-layer dynamical activity.

\begin{figure}
    \centering
    \includegraphics[width=0.94\textwidth]{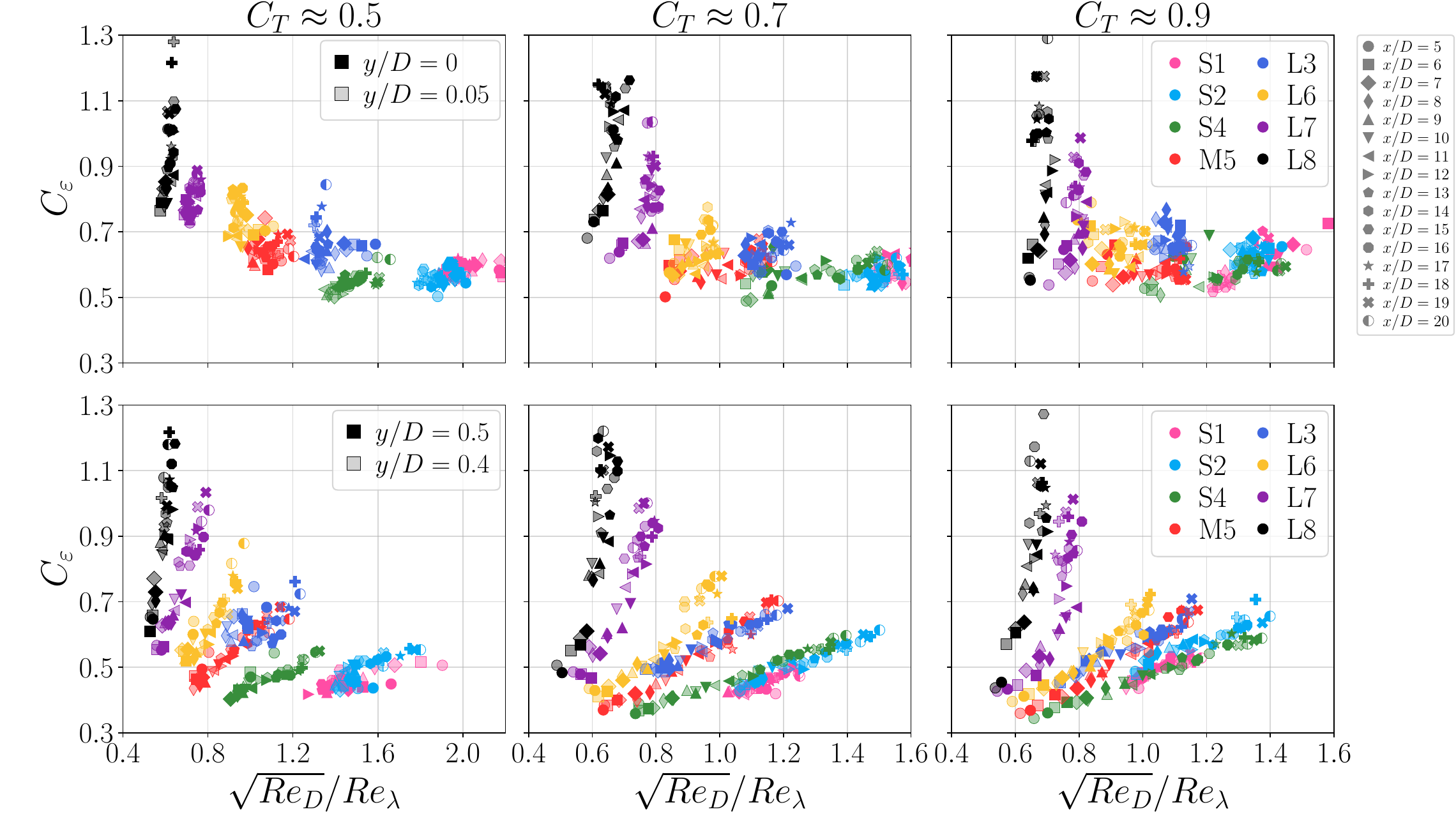}
    \caption{Normalised dissipation coefficient $C_{\varepsilon}(x)$ plotted against $\sqrt{Re_D}/Re_{\lambda}(x)$ for the various flavours of FST and values of $C_T$ examined. Top row: measurements along ($y/D=0$, filled symbols) and adjacent  to ($y/D=0.05$, open symbols) the wake centreline. Bottom row: measurements in the blade-tip region  ($y/D=0.5$, filled symbols; $y/D=0.4$, open symbols). Colours denote the different FST cases, and symbol shapes indicate the streamwise location within the wake. }
    \label{fig:Ceps_Ray}
\end{figure}

The normalised dissipation coefficient fields reveal additional interesting features (figure~\ref{fig:Ceps_field}). For the low- and moderate-turbulence-intensity inflows (S1 to L6), a ring of elevated $C_{\varepsilon}$ surrounds the wake, demarcating the two adjacent streams of turbulence: that in the free stream (produced by the active turbulence-generating grid) and that in the wake (generated by the mean shear). The ring becomes both wider and more intense with increasing $C_T$. As $I_{\infty}$ increases, however, the coherence of this region of elevated $C_{\varepsilon}$ progressively weakens and the ring ultimately vanishes at the highest FST intensities. A further observation is that $C_{\varepsilon}$ in the free stream is typically larger than within the wake, and the two decay at different rates owing to the different initial conditions of the two turbulent streams.

We now turn to the streamwise evolution of $C_{\varepsilon}(x)$ along the wake centreline and in the blade-tip region (figure~\ref{fig:Ceps_stream}). In the very near field ($x/D\lesssim 5$), $C_{\varepsilon}$ varies in the opposite direction to $Re_{\lambda}$; decreasing whilst $Re_{\lambda}$ increases. Subsequently, for low- and medium-turbulence-intensity inflows (S1 to L6), it can be observed that $C_{\varepsilon} \approx \textrm{const.}$ along the wake centreline ($y/D=0$) for $x/D \gtrsim 5$, over approximately the same streamwise range where $Re_{\lambda}(x)$ remains nearly constant. By contrast, in the outer wake region ($y/D=0.5$), where $Re_{\lambda}(x)$ was found to decrease, $C_{\varepsilon}(x)$ increases with streamwise distance, consistent with the inverse relationship reported by \citet{Apostolidis_Laval_Vassilicos_2022,Chen_Cuvier_Foucaut_Ostovan_Vassilicos_2021,Noriega_2025}. A similar inverse behaviour is observed in the horizontal direction, with $C_{\varepsilon}(x=\textrm{const.},y)$ increasing in regions where $Re_{\lambda}(x=\textrm{const.},y)$ decreases and \emph{vice versa} (horizontal profiles are not shown here for brevity). 

The picture is more complex for the two highest turbulence-intensity inflows (L7 \& L8). Along the wake centreline, $C_{\varepsilon}\approx\textrm{const.}$ for L7 for $x/D\ge 5$ but only a very limited quasi-constant interval can be observed for L8. At higher $C_T$, however, no clearly identifiable region of $C_{\varepsilon}\approx\textrm{const.}$ is observed, with $C_{\varepsilon}$ instead increasing with streamwise distance. In the outer wake ($y/D=0.5$), $C_{\varepsilon}$ similarly increases downstream for $x/D\ge5$, where $Re_{\lambda}$ decreases, consistent with the inverse $C_{\varepsilon} - Re_{\lambda}$ relationship. 

The assessment of the equilibrium/non-equilibrium scaling of the dissipation rate follows naturally from these observations. Figure~\ref{fig:Ceps_Ray} presents scatter plots of the normalised dissipation coefficient, $C_{\varepsilon}(x)$, for the different thrust coefficients and flavours of FST as a function of $\sqrt{Re_D}/Re_{\lambda}(x)$. Marker colours have been changed in these plots to clearly distinguish the different turbulent inflows. The wake-core ($y/D \in \{0,0.05\}$) and blade-tip regions ($y/D \in \{0.4,0.5\}$) are shown separately, with only data for $x/D \geq 5$ included. In the central wake region (upper panel of figure~\ref{fig:Ceps_Ray}), we observe that $C_{\varepsilon}\approx\textrm{const.}$ for low- and medium-turbulence-intensity inflows, in particular for the lowest $C_T$ values, which is indicative of either classical equilibrium dissipation or balanced non-equilibrium \citep{Goto2016b}. The degree of scatter on the $C_{\varepsilon}$ data is slightly amplified by increasing the thrust coefficient to $C_T\approx 0.9$,  yet the relationship $C_{\varepsilon}\approx\textrm{const.}$ still holds reasonably well for low- to moderate-$I_{\infty}$ inflows. The magnitude of $C_{\varepsilon}$ across the FST flavours S1 to M5 is also relatively insensitive to $C_T$ (across all three thrust coefficients). Hence, although increasing the forcing amplitude ($C_T$) intensifies the wake turbulence relative to the background turbulence, it does not seem to modify the efficiency of the energy cascade ($C_{\varepsilon}$).

For the inflows with the highest $I_{\infty}$ (L7 \& L8), there is a very large scatter of $C_{\varepsilon}$ but this occurs at almost constant $\sqrt{Re_D}/Re_{\lambda}$ throughout the wake's evolution, commensurate with the $Re_{\lambda}$ plateaus shown in figure~\ref{fig:Reynolds_Taylor}. The dissipation coefficient can, to some extent, be interpreted as the efficiency of the dissipation with respect to the inertial inter-scale cascade of TKE. This implies that for the most turbulent inflows, whilst the wake turbulence decays only weakly ($Re_{\lambda}(x) \approx \textrm{const.}$), the efficiency of the dissipation increases, presumably as a result of the enhanced mixing of the two streams of turbulence compared to the weakly turbulent inflows. Consistent with this interpretation, $C_{\varepsilon}$ increases both with increasing FST and with decreasing $C_T$, i.e. whenever the ratio of the wake-turbulence to FST decreases, suggestive of the fact that the interaction between two similar streams of turbulence somehow enhances the efficiency of the dissipation. We further examine this enhanced dissipation efficiency in light of the local flow intermittency in the next section. 


The picture is much clearer in the outer portion of the wake. For low- and moderate-turbulence-intensity inflows and all $C_T$ values, there is a clear trend of $C_{\varepsilon}\sim \sqrt{Re_D}/Re_{\lambda}$, providing evidence of non-equilibrium dissipation scaling in the blade-tip region. On the other hand, for the FST flavours with the highest $I_\infty$ (L7 \& L8), $C_{\varepsilon}$ does not appear to scale linearly with $\sqrt{Re_D} / Re_\lambda$, especially at the highest $C_T$. Higher-order non-equilibrium scaling exponents were also tested, notably $C_{\varepsilon} \sim Re_D/Re_{\lambda}^2$ ($n=m=2$), but these did not significantly improve on the $n=m=1$ scaling. In these cases, wake-generated and ambient turbulence rapidly become comparable in intensity, and no single exponent pair appears to provide an exact scaling. Given this interaction between two comparable streams of turbulence, it is also plausible that a scaling based on a single turbulent Reynolds number is inadequate for capturing the nature of the cascade/dissipation. Indeed, the dissipation-scaling framework was originally developed for turbulence decaying into a non-turbulent background, i.e. with a single turbulence generator driving the energy injection into the cascade. In the highly turbulent FST cases, however, neither the wake-generated nor the ambient turbulence is markedly stronger than the other, so this assumption of a single injection mechanism may no longer hold. Moreover, as a locally aggregated quantity, $Re_\lambda$ does not distinguish between the wake-generated and ambient contributions to the local turbulence :  the two, despite being of comparable intensity, are of different origin and may correspond to two distinct turbulent cascades that cannot be resolved with a single turbulent Reynolds number.


\section{Wake intermittency \label{sec:wake intermittency}}

Turbulent flows are inherently intermittent at small scales : the heavy tails of the velocity-increment statistics within the dissipative and inertial ranges reflect rare, localised events of intense energy dissipation \citep{Kolmogorov_1962,Meneveau_Sreenivasan_1991,Elsinga2020}. A distinctive feature of porous-disc and wind-turbine wakes, however, is that intermittency also arises at large scales : as outlined in \S~\ref{sec:Introduction}, a growing body of evidence shows that these wakes are surrounded by a ring of highly intermittent flow at scales comparable to the rotor diameter \citep[e.g.][]{Schottler2018,Neunaber2020,Vinnes2023}. In this section, we examine the influence of FST and $C_T$ on the intermittency ring, and its connection with the dissipation characteristics discussed previously.

Intermittency is characterised across scales through the statistics of the longitudinal velocity increments. At each measurement station, temporal increments are computed from the hot-wire time series as $\delta u(t,\tau) = u(t+\tau) - u(t)$, and converted into spatial increments, $\delta u(x,\ell) = u(x+\ell) - u(x)$, using Taylor’s frozen-turbulence hypothesis with $\ell = \tau U(x,y)$. Two approaches are classically employed to investigate intermittency: the scaling of the structure functions---the moments $\langle \delta u^{n} \rangle $ of the velocity increments---and the deviation of the probability density function $p(\delta u)$ from Gaussianity for different separations $\ell$. The departure from statistical self-similarity of the increments across scales is quantified by the scale-dependent shape parameter $\Lambda^2(\ell) = \textrm{ln}(F(\delta u(x,\ell))/3)/4$ where $F(\delta u(x,\ell))$ is the flatness ($\sim \langle \delta u^4 \rangle$) of the velocity increments \citep{Chilla1996}. In Kolmogorov’s 1962 theory, $\Lambda^2(\ell)$ is related to the variance of the logarithm of the energy dissipation rate locally averaged at scale $\ell$, thereby providing a proxy for characterising intermittent dissipation events across scales \citep{Kolmogorov_1962,CASTAING1990,Schmitt2025}. Here, we will use this second approach. 

\begin{figure}
    \centering
    \includegraphics[width=\linewidth]{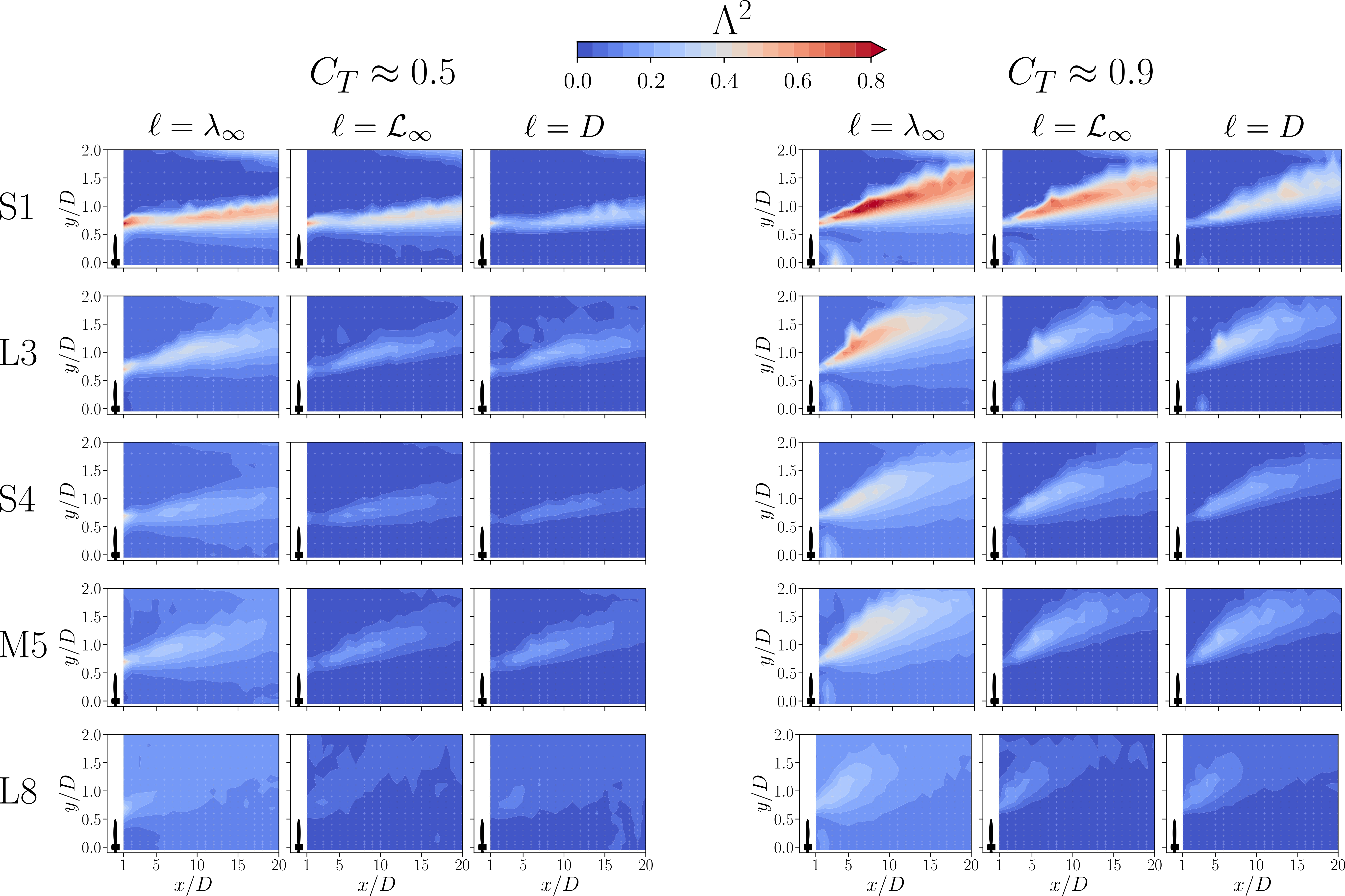}
    \caption{Spatial maps of the shape factor $\Lambda^2$ at three different separation scales $\ell \in \{ \lambda_{\infty}, {\cal L}_{\infty}, D \}$ for five inflow conditions and $C_T \in \{0.5,0.9\}$.}
    \label{fig:lambda_maps}
\end{figure}

Figure~\ref{fig:lambda_maps} presents spatial maps of the shape parameter $\Lambda^2$ at three separation scales : the free stream Taylor microscale $\ell=\lambda_{\infty}$ and integral length scale $\ell = {\cal L}_{\infty}$ (both specific to each inflow), and the rotor diameter $\ell = D$, for five inflow conditions and two thrust coefficients. 
In all cases, intermittency is concentrated in an annular region emanating from the rotor tip and spreading radially outwards with downstream distance : this is the intermittency ring introduced in \S~\ref{sec:Introduction}, here mapped throughout the wake and across turbulent scales.
Flow intermittency is most pronounced at small scales ($\ell = \lambda_\infty$), yet persists at large scales ($\ell \geq D$) especially for weakly turbulent background flows (e.g. S1). Increasing the thrust coefficient intensifies and widens the ring for every inflow, consistent with the stronger and more dynamically active shear layer at higher rotor loading, whereas increasing FST intensity progressively erodes it, especially at the large scales. In particular, whilst the flow at the wake edge is intermittent at both  rotor, injection and dissipative scales for the least turbulence case (S1), only small-scale intermittency is observed for the most turbulent inflow (L8). This attenuation seems to be governed primarily by the presence of FST intensity $I_{\infty}$, with a weak influence of the turbulence integral length scale ${\cal L}_{\infty}$. 

\begin{figure}
    \begin{subfigure}[c]{\linewidth}
        \emph{(a)} \par \vspace{0.1cm}
        \centering
        \includegraphics[height=5cm]{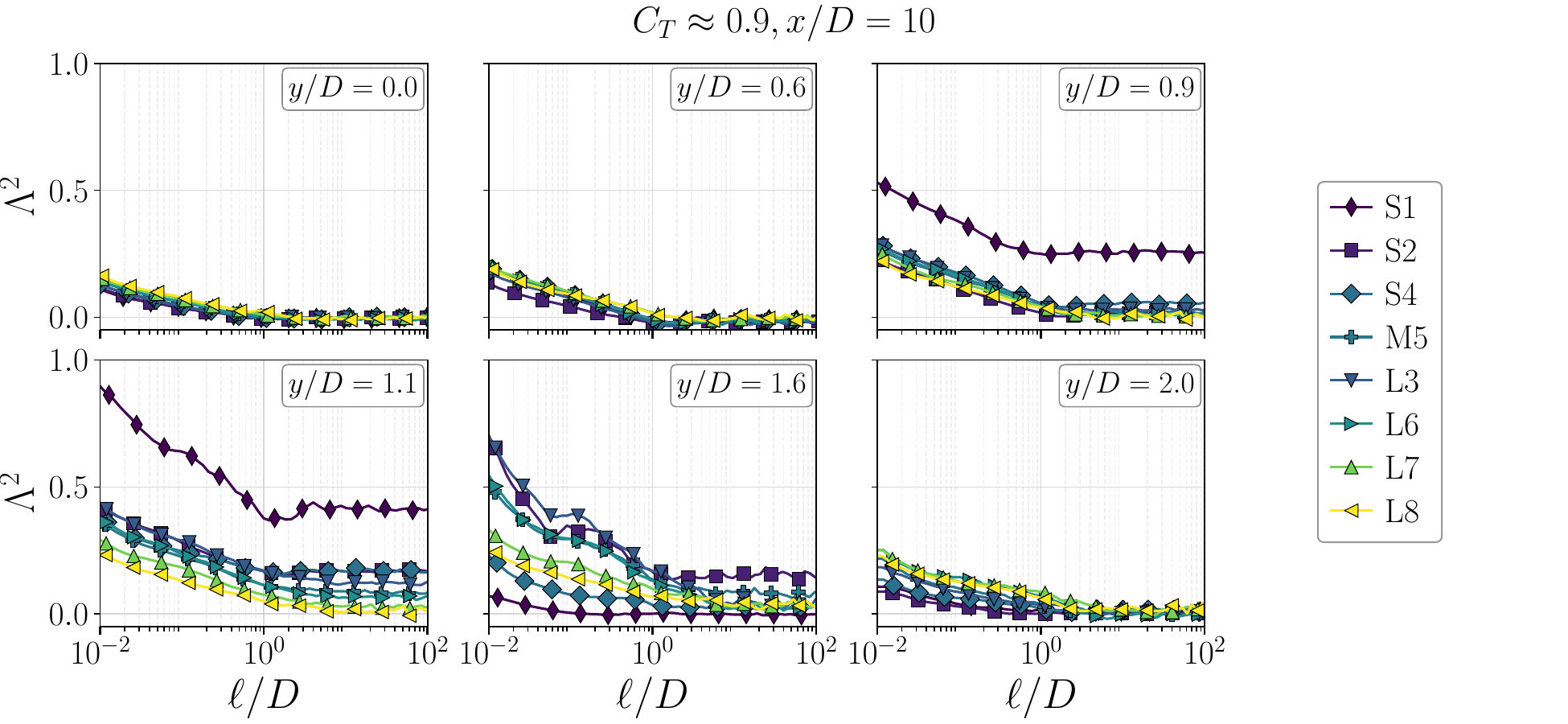}
    \end{subfigure}
    \\
    \begin{subfigure}[c]{0.9\linewidth}        
        \emph{(b)} \par \vspace{0.1cm}
        \centering
        \includegraphics[width=\linewidth]{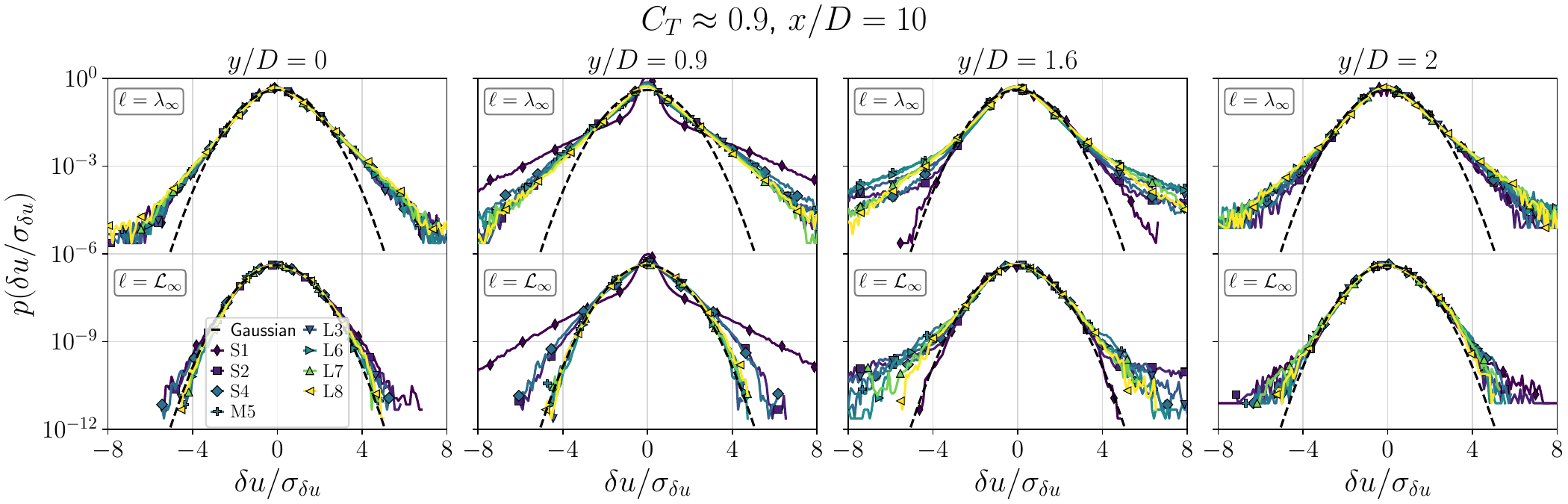} \\ \vspace{0.5em}
        \includegraphics[width=\linewidth]{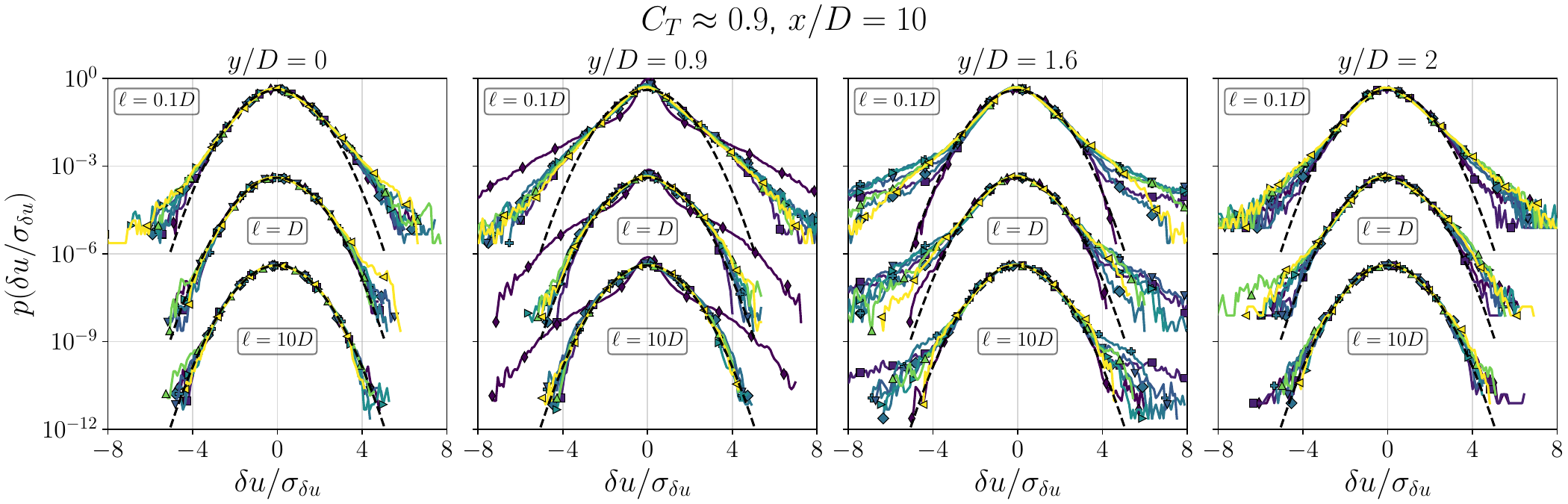}
    \end{subfigure}     
     \begin{subfigure}[c]{0.9\linewidth}        
        \emph{(c)} \par \vspace{0.1cm}
        \centering
        \includegraphics[height=4cm]{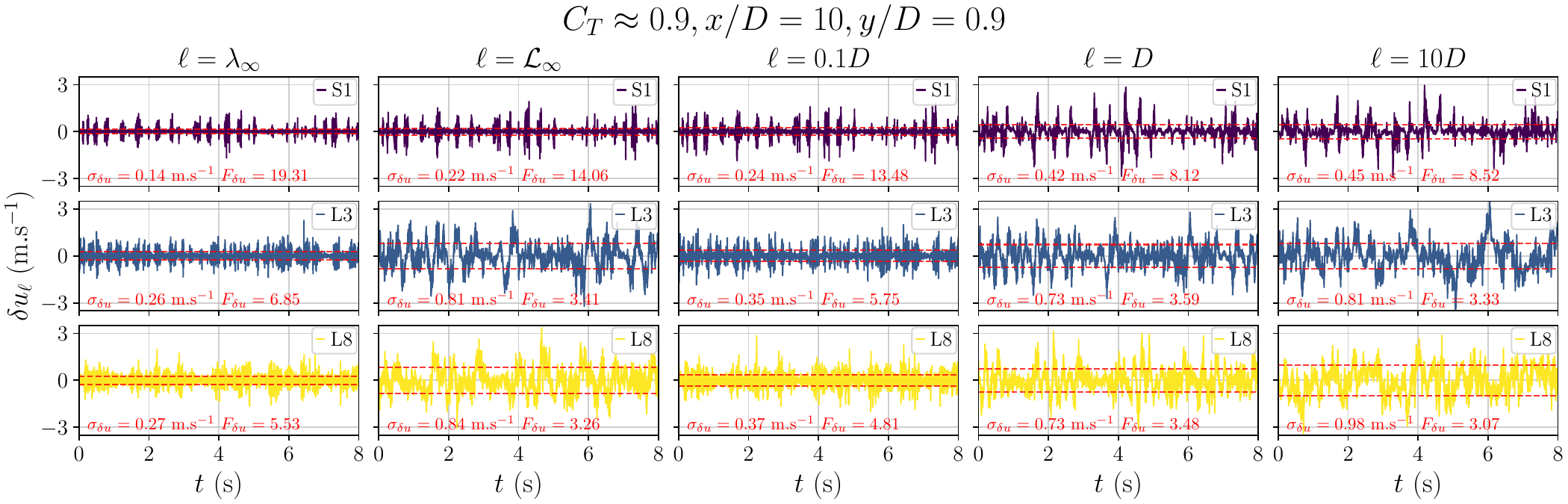}
        
        \end{subfigure}
    \caption{Intermittency of the streamwise velocity increments $\delta u(\ell)$
    at $x/D = 10$ for $C_T \approx 0.9$, for all inflow conditions (S1--L8, see legend).
    (\emph{a}) Shape factor as a function of the separation length $\ell/D$ at six transverse locations spanning the inner wake, the shear layer and the free stream. 
    (\emph{b}) Normalised p.d.f of $\delta u$
    for varying separation lengths $\ell/D$ based on the turbulence length scales (top) and on the turbine diameter (bottom); dashed lines denote a Gaussian distribution and the p.d.f are shifted vertically for clarity. 
    (\emph{c}) Corresponding time series of $\delta u$ at $y/D = 0.9$ for the inflows S1, L3 and L8 (rows) and the five separation lengths listed above (columns); the values indicated in red are the standard deviation $\sigma_{\delta u}$ and the flatness $F_{\delta u}$.}
    \label{fig:lambda_profiles}
\end{figure}

The structure of this ring is examined in more detail in figure~\ref{fig:lambda_profiles}, for $C_T \approx 0.9$ at $x/D = 10$. The profiles of $\Lambda^2(\ell)$ (figure~\ref{fig:lambda_profiles}(a)) quantify the contrast between the wake core and its edge. At the centreline, $\Lambda^2(\ell)$ is significant only at the smallest scales and decays rapidly with \(\ell\) for all inflows : the core exhibits the classical small-scale intermittency of homogeneous isotropic turbulence. Within the ring $(y/D \in [0.9,1.6])$, by contrast, the inflow conditions differ strongly. For the least turbulent cases, $\Lambda^2(\ell)$ remains elevated across all scales, including \(\ell \gg D\), whereas increasing FST intensity progressively suppresses large-scale intermittency. The corresponding increment PDFs (figure~\ref{fig:lambda_profiles}(b)) confirm this observation : close to Gaussian at large separations on the centreline, they develop pronounced heavy tails at all separations---up to \(\ell = 10D\)---at \(y/D = 0.9\) for the weakly turbulent inflows. The velocity time series at this position (figure~\ref{fig:lambda_profiles}(c)) reveal the origin of this large-scale non-Gaussianity. For case S1, the increment signal alternates between quiescent periods and intense, clustered bursts, presumably associated with the meandering and flapping of the tip shear layer driven by the turbine-generated coherent structures, which persist under low FST intensity \citep{Bourhis2025}. By contrast, for the most turbulent case, L8, the signal is uniformly agitated and this on-off character disappears.

The dissipation behaviour discussed previously can then be interpreted in light of these intermittency characteristics. Indeed, cascade intermittency reflects local imbalances between inter-scale energy flux and dissipation, which may in turn contribute to variations in $C_{\varepsilon}$ \citep{Cafiero2019}. First, for mild to moderate FST intensities, the ring of enhanced intermittency---spanning both small ($\ell \leq \lambda$) and large ($\ell\geq D$) scales---coincides with the ring of elevated $C_{\varepsilon}$ reported above. This increased intermittency implies stronger ``dissipation hot spots" (i.e. localised regions of exceptionally high dissipation magnitude) which may collectively increase the mean dissipation rate, and hence $C_{\varepsilon}$. In addition, a plausible origin for the non-equilibrium dissipation found in the outer wake may lie in this large-scale intermittency. Extreme events at the energy-containing scales raise the spectral content at low wavenumbers, thereby perturbing the cascade's ``energy-containing reservoir". Since the perturbation occurs at the large scales, a finite time is required for it to cascade down to the dissipative range, so that the dissipation rate no longer adjusts instantaneously to the rate of energy injection to the inertial range; inter-scale flux and dissipation are then out of balance, which is the defining condition of non-equilibrium dissipation. The particular type of non-equilibrium dissipation scaling $C_{\varepsilon} \sim \sqrt{Re_D}/Re_\lambda$ ($m=n=1$) found in the outer wake, for low- to moderate-intensity FST, may therefore result from this large-scale intermittency. By contrast, intermittency in the wake core is only weakly enhanced and remains confined to the small scales. The wake core therefore exhibits characteristics of homogeneous isotropic turbulence, with $C_{\varepsilon}$ remaining approximately constant and intermittency restricted to the small scales, consistent with the refined K62 similarity hypothesis \citep{Kolmogorov_1962} and previous observations in wind-turbine wakes \citep{Neunaber2020}.

For the most intense turbulent inflows, intermittency in the outer wake is enhanced only at the small scales, and no clear scaling for the dissipation rate could be identified in either the inner or the outer wake. The absence of large-scale intermittency is consistent with the early suppression of the turbine-generated structures; with the tip vortices broken down close to the rotor, one particular (efficient) pathway to perturbing the cascade through the low-wavenumber energy-containing scales is broken. Consequently, any ``kick''/perturbation may instead be introduced closer (in wavenumber space) to the dissipative range (small-scale intermittency). Accordingly, the time lag associated with energy cascading to the dissipation range is shorter, and the imbalance between the inertial inter-scale flux and the dissipation rate correspondingly weaker, which may induce a different non-equilibrium regime.

\section{Conclusions and wider discussion}

Wind-tunnel experiments were conducted using a $0.58$~m diameter model wind turbine operating at three thrust coefficients and exposed to a range of FST conditions generated by an active turbulence-generating grid.

We observed that increasing the FST intensity accelerates the recovery of the mean velocity deficit while simultaneously increasing the turbine-added turbulence intensity in the near field. For most turbulent inflows, however, no appreciable difference in turbulence intensity between the wake and the free stream persists in the far field, as a result of enhanced turbulent transport and mixing. At low and moderate FST intensities, the integral length scale in the outer wake grows with downstream distance : the tip vortices retain their coherence over several diameters and organise into progressively larger energy-containing structures, which induce the flapping and meandering of the tip shear layer \citep{Bourhis2025,Biswas_Buxton_2026}. As the FST intensity increases, this coherence is destroyed within the first few rotor diameters, before vortex pairing can occur : the tip shear layer therefore ceases to act as an autonomous source of turbine-generated structures, and the outer-wake integral length scale progressively adjusts to that of the free stream.

We then examined the self-similar evolution of the wake. For all inflow conditions, the mean velocity-deficit profiles attained self-similarity relatively close to the turbine ($x/D\approx3$). Under weak FST intensity the collapse persisted to the farthest downstream measurement station, whereas at high FST intensity it deteriorated in the far field, where the residual deficit became comparable to the background velocity fluctuations. The streamwise normal Reynolds-stress profiles followed the same trend, remaining self-similar throughout the measurement domain at low- and moderate-intensity FST but losing self-similarity downstream in high-intensity FST, where the wake and free-stream turbulence intensities became comparable. In the near wake, the dissipation rate of turbulent kinetic energy peaked in the blade-tip region, where the mean shear is strongest; farther downstream the maximum moved inwards, following the spread of the tip shear layer towards the wake centreline. The dissipation-rate profiles also exhibited approximate self-similarity, although the quality of the collapse degraded with increasing FST intensity. Overall, increasing FST intensity promoted an earlier onset of self-similar behaviour but also led to its earlier breakdown in the far field, as also reported for bluff bodies by \citet{Rind2012}.

We then examined the evolution of the Taylor Reynolds number, $Re_{\lambda}$, and of the normalised dissipation rate, $C_{\varepsilon}$. In the near field $Re_{\lambda}$ increased, with a maximum in the tip shear layer where the turbine-generated structures enhance the scale separation. Farther downstream its behaviour depended on radial position : $Re_{\lambda} $ remained approximately constant along the centreline but spatially decayed in the outer wake.
At low and moderate FST intensities, a ring of elevated $C_{\varepsilon}$ formed at the periphery of the wake, encircling the regions of mean velocity deficit and turbine-added turbulence intensity. The ring widened and intensified with $C_T$, but was progressively eroded as the FST intensity increased. In general, $Re_{\lambda}$ and $C_{\varepsilon}$ were observed to vary inversely across the wake, the one decreasing where the other increased. These observations motivated a closer examination of the dissipation scaling. For weak and mild inflow turbulence intensities, $C_{\varepsilon}$ remained approximately constant in the far field, indicative of either classical equilibrium scaling or a balanced non-equilibrium state \citep{Goto2016b}. However, in the outer region of the wake there was a clear scaling, $C_{\varepsilon}\sim\sqrt{Re_D}/Re_{\lambda}$ which is indicative of a particular type of non-equilibrium dissipation scaling corresponding to $C_{\varepsilon}~\sim Re_D^{m/2}/Re_{\lambda}^n$ with $n=m=1$. For the most turbulent inflows, $C_{\varepsilon}$ varied substantially with streamwise distance and no comparable scaling could be identified using a single turbulent Reynolds number.

The scale-dependent distribution of intermittency offered a plausible interpretation of these changes in the dissipation regime. For weakly turbulent inflows, the annular region of elevated $C_{\varepsilon}$ was found to coincide spatially with a ring of enhanced intermittency at both small ($\ell \leq \lambda_{\infty}$) and large ($\ell \geq D$) scales. The enhanced small-scale intermittency indicates a greater occurrence of intense dissipation events, which may contribute collectively to increase the mean dissipation rate. The concurrent enhancement of large-scale intermittency provides a possible physical connection with the observed non-equilibrium dissipation regime. In the blade-tip region, large-scale intermittency is driven by persistent, energetic coherent motions associated with the turbine-generated tip shear layer and tip-vortex system. Such dynamics can produce a strongly unsteady supply of energy to the cascade at low wavenumbers. Since the transfer of energy to the dissipative scales occurs over a finite cascade time, ``kicks" introduced at the large scales are not transmitted instantaneously to the small scales. The inter-scale energy flux and dissipation may therefore cease to be in local equilibrium, providing a plausible mechanism for the non-equilibrium dissipation scaling observed in the outer wake under mild- and moderate-intensity FST conditions. At the wake centreline, by contrast, intermittency was only weakly enhanced at small scales with the dissipation rate following the equilibrium scaling. These results further suggest a connection between the nature of the turbulence dissipation and the scale-dependent distribution of intermittency, as also outlined by \citet{Schmitt2025}.

For highly turbulent inflows, intermittency in the outer wake was only marginally enhanced at small scales, with no comparable enhancement at large scales. This behaviour is consistent with the accelerated breakdown and reduced coherence of turbine-generated tip-vortex dynamics under elevated FST intensity, thereby suppressing large-scale intermittent perturbations to the cascade. Intermittent events therefore occur predominantly at higher wavenumbers, closer to the dissipative range (in wavenumber space), potentially shortening the cascade time and reducing the imbalance between inter-scale energy flux and dissipation. This distinct intermittency regime may correspond to a different non-equilibrium dissipation regime with a different pair of exponents $\{m,n\}$. Alternatively, describing the turbulence with a single turbulent Reynolds number ($C_{\varepsilon}~\sim Re_D^{m/2}/Re_{\lambda}^n$) may also be insufficient to describe the mixing of two comparable streams of turbulence where one does not dominate the other, as is seen in the other cases (or indeed when the background is non-turbulent; the situation in which this non-equilibrium dissipation scaling was first discovered).

Future work should examine in greater detail the vicinity of the turbulent/turbulent interface (TTI) separating the wake from the surrounding FST, with the aim of relating the interfacial dynamics, and the vortical structures populating it, to the ring of elevated $C_{\varepsilon}$ and enhanced intermittency. A natural continuation of this work would be to address, from a more fundamental standpoint, the phenomenology of the cascade in the presence of large-scale intermittent events, and in particular the time lag between between energy injection bursts at the large scales and the dissipation at the small scales. Finally, the dissipation scaling remains to be established for two adjacent turbulent streams of comparable intensity, where small-scale intermittency alone exists in the outer wake region. Taken together, these questions point towards a broader one: whether a universal relationship exists between the intermittency of the cascade and the dissipation scaling.

\begin{bmhead}[Acknowledgements]
The authors would like to thank Michael Hölling for providing access to the facility, Lars Neuhaus, Julian Jüchter, Felix Schmidt, and Thomas Messmer for their assistance with the experiments and for many fruitful discussions, and John Christos Vassilicos for his feedback on an early draft of this paper.
\end{bmhead}

\begin{bmhead}[Funding]
This work was supported by the Engineering and Physical Sciences Research Council (grant number EP/V006436/1).
\end{bmhead}

\begin{bmhead}[Declaration of interests]
The authors report no conflict of interest. 
\end{bmhead}

\begin{bmhead}[Data availability statement]
Data are available upon request to the corresponding author. For the purposes of open access, the authors have applied a Creative Commons Attribution (CC BY) licence to any Author Accepted Manuscript (AAM) version arising.
\end{bmhead}

\begin{bmhead}[Author ORCIDs]
Martin Bourhis, https://orcid.org/0000-0003-3107-032X; Oliver R.H. Buxton, https://orcid.org/0000-0002-8997-2986.
\end{bmhead}

\bibliographystyle{jfm}
\bibliography{jfm}

\begin{appen}

\section{Active turbulence-generating grid protocols} \label{appA}

\normalsize

This appendix provides further details on the active-grid protocols used to generate the different FST ``flavours''. Table~\ref{tab:grid_protocols} reports, for each turbulent inflow condition, the time-averaged absolute flap angle over a grid cycle, $\langle \lvert\alpha(t)\rvert \rangle$; the standard deviation of the flap-angle fluctuations, $\sigma(\alpha)$; and the mean and standard deviation of the shaft rotation rate, $\Omega$ and $\sigma(\Omega)$, respectively. Except for the static cases S1 and S4, $\langle |\alpha(t)| \rangle$ is identical for all cases, i.e. the total grid blockage of the flow is fixed across protocols, both instantaneously and when averaged over a complete grid cycle. The protocols differ in the amplitude of the flap-angle fluctuations and in both the mean and standard deviation of the shaft rotation rate. A schematic of the active grid, together with example shaft-angle time series for five protocols and their corresponding power spectral densities, is provided in figure~\ref{fig:appendix}. Further details of the active-grid design and capabilities can be found in \citet{Lars2020,Neuhaus2021,PhD_Lars}. Movies of the grid in operation are provided in \citet{Bourhis2025}.

\begin{table}
\centering
\setlength{\tabcolsep}{7pt} 
\def~{\hphantom{0}}
  \begin{tabular}{lccccccccccc}
       \multicolumn{2}{c}{FST Case} &  Mode  &  $\langle {\lvert\alpha(t)\rvert} \rangle~(\textrm{deg})$ & $\sigma(\alpha)~(\textrm{deg})$ & $\Omega~(\textrm{deg}.\textrm{s}^{-1})$ & $\sigma(\Omega)~(\textrm{deg}.\textrm{s}^{-1})$    \\[3pt]
       S1 & \Sone   & SM & 0    & -   & - & -\\
       S2 & \Stwo   & SM & 30   & -   & - & -\\
       S4 & \Sfour  & SM & 0--90& -   & - & - \\
       M5 & \Mfive  & DM &  30  & 10  & 152 & 108 \\
       L3 & \Lthree & DM &  30  & 2.5 & 10 & 9\\
       L6 & \Lsix   & DM &  30  & 5   & 20 & 17\\
       L7 & \Lseven & DM &  30  & 10  & 39 &  34\\
       L8 & \Leight & DM &  30  & 15  & 61 &53\\    
  \end{tabular}
  \caption{Active-grid protocols used to generate the free-stream turbulence conditions. SM and DM denote the static and dynamic operating modes, respectively. $\langle \lvert\alpha(t)\rvert \rangle$ is the time-averaged absolute flap angle over a grid cycle about the vertical or horizontal axes (see figure~\ref{fig:appendix}(\textit{a})); $\langle \lvert\alpha(t)\rvert \rangle = 0^\circ$ corresponds to flaps parallel to the flow (fully open), and $\langle \lvert\alpha(t)\rvert \rangle = 90^\circ$ to flaps normal to the flow (fully closed). $\sigma(\alpha)$ is the standard deviation of the flap-angle fluctuations, and $\Omega$ and $\sigma(\Omega)$ are the mean and standard deviation of the axis rotation rate. Further details of the protocols are given in \citet{Bourhis2025}.}
  \label{tab:grid_protocols}
\end{table}

\begin{figure}
    \begin{subfigure}[c]{0.45\linewidth}
        \emph{(a)} \par \vspace{0.1cm}
        \centering
        \includegraphics[width=\linewidth]{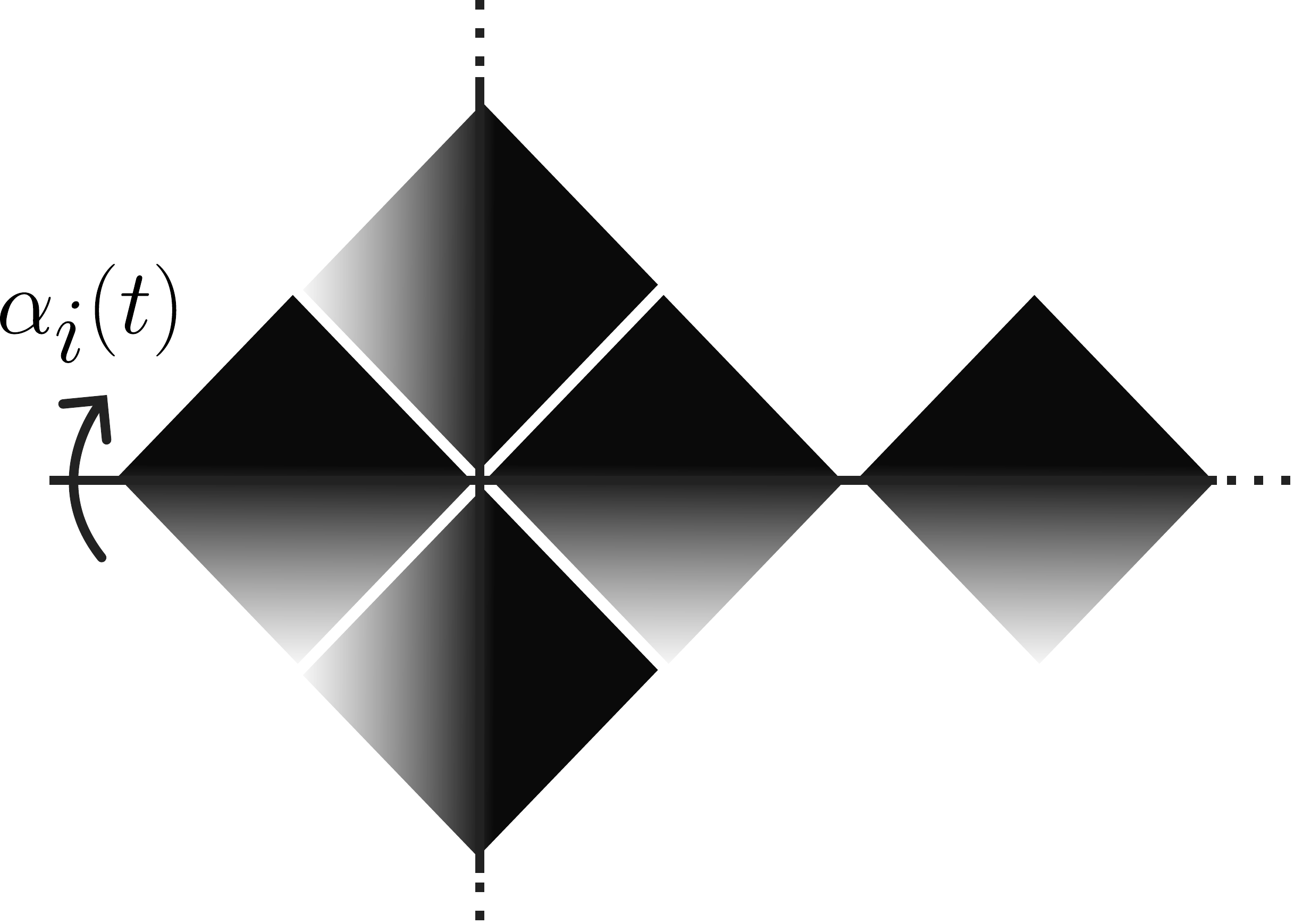}
    \end{subfigure}
    \hfill
    \begin{subfigure}[c]{0.45\linewidth}        
        \emph{(b)} \par \vspace{0.1cm}
        \centering
        \includegraphics[width=\linewidth]{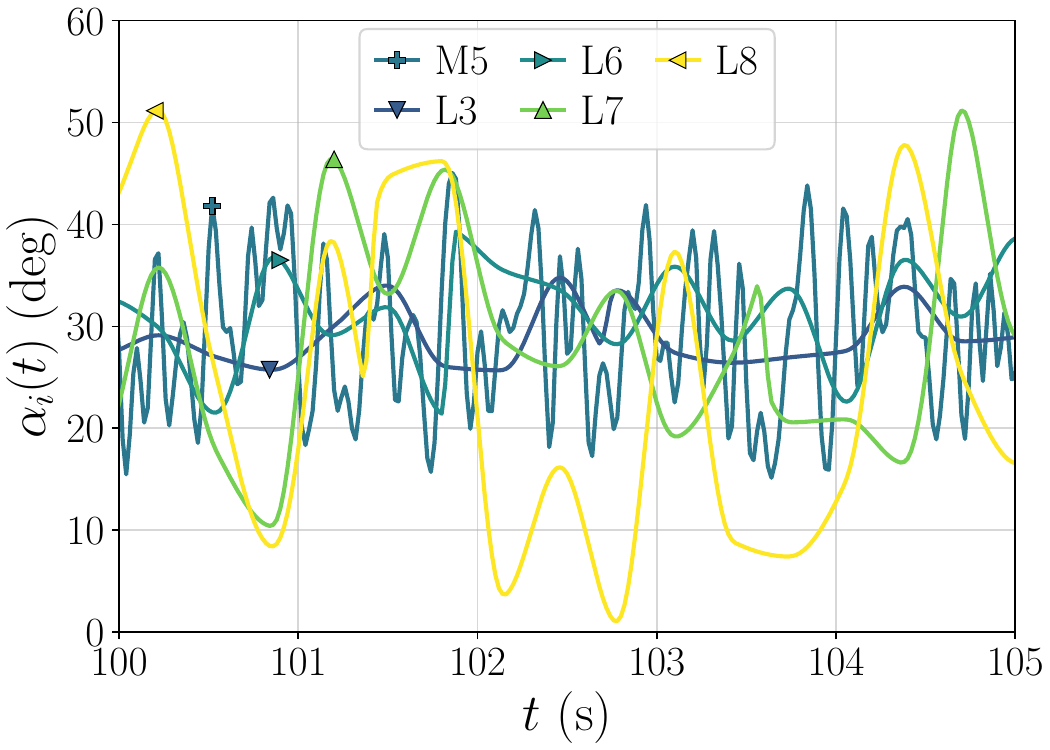}
    \end{subfigure}
    \\
        \centering
        \begin{subfigure}[c]{0.45\linewidth}        
        \emph{(c)} \par \vspace{0.1cm}
        \centering
        \includegraphics[width=\linewidth]{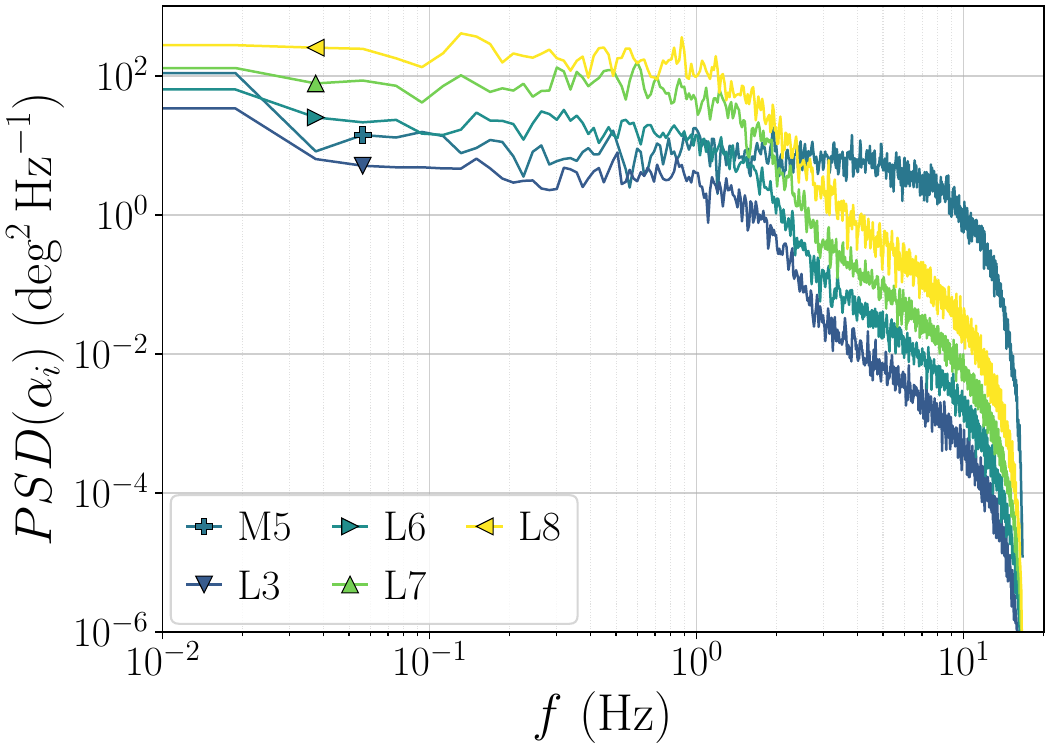}
    \end{subfigure}
     \hfill
    \caption{(\textit{a})~Schematic of the active grid; $\alpha_i(t)$ denotes the instantaneous angle of shaft $i$. (\textit{b})~Sample shaft-angle time series $\alpha_i(t)$ for the 
five dynamic-mode (DM) actuation protocols (see table~\ref{tab:grid_protocols}). (\textit{c})~Power spectral densities of the corresponding shaft-angle signals.}
    \label{fig:appendix}
\end{figure}


\end{appen}\clearpage

\end{document}